\documentclass[10pt]{article}
\usepackage{amsmath}
\usepackage{amssymb}
\usepackage{amsfonts}
\usepackage{graphicx}

\allowdisplaybreaks
\input{tcilatex}

\begin{document}

\begin{center}
{\Large Nonlinear continuous integral-derivative observer}

Xinhua Wang and Bijan Shirinzadeh

{\small \ Department of Mechanical and Aerospace Engineering,}

{\small Monash University, Melbourne, VIC, 3800, Australia Email:
wangxinhua04@gmail.com Tel: +61 0432 704528}
\end{center}

\textbf{Abstract.} In this paper, a high-order nonlinear continuous
integral-derivative observer is presented based on finite-time stability and
singular perturbation technique. The proposed integral-derivative observer
can not only obtain the multiple integrals of a signal, but can also
estimate the derivatives. Conditions are given ensuring finite-time
stability for the presented integral-derivative observer, and the stability
and robustness in time domain are analysed. The merits of the presented
integral-derivative observer include its synchronous estimation of integrals
and derivatives, finite-time stability, ease of parameters selection,
sufficient stochastic noises rejection and almost no drift phenomenon. The
theoretical results are confirmed by computational analysis and simulations.


\bigskip

\textbf{1. Introduction}

Time derivatives and integrals are important components in almost all
engineering applications. The problems of the time derivatives and integrals
are those of estimating the values $D_{i}\left( a\right) =\frac{d^{i}a\left(
t\right) }{dt^{i}}$ and $I\left( a\right) =\int_{0}^{t}\cdots
\int_{0}^{s}a\left( \sigma \right) d\sigma \cdots d\tau $, respectively.
Obtaining the positions, velocities and accelerations is crucial for many
types of systems for correct and timely performances, such as missile flying
control systems [1] and the antilock braking systems [2]. For an airplane,
the flying velocity can be obtained from an airspeed indicator. In order to
design a feedback controller, the time integral and derivative performances
are required to estimate the position and acceleration trajectories of the
airplane, respectively. For the guidance of a missile system, the
acceleration is obtained from the accelerometer. In order to hit the target,
the position trajectory of the missile should be obtained. Therefore, in the
circumstances without GPS, we need to compose a double-integral algorithm to
estimate the position trajectory from the acceleration measurement on
condition that the initial position and velocity are known. Inertial
Navigation System (INS) is a self-contained navigation technique in which
measurements provided by accelerometers and gyroscopes are used to track the
position and orientation of an object relative to a known starting point,
orientation and velocity. In INS, Inertial measurement units (IMUs)
typically contain three orthogonal rate-gyroscopes and three orthogonal
accelerometers, measuring angular velocity and linear acceleration
respectively. To calculate the position of the device the signals from the
accelerometers are double integrated. However, for a long-time navigation,
the drift phenomenon of INS is mainly brought out by the usual integral
methods. They cannot restrain the effect of stochastic noise, especially
non-white noise. Such noise leads to the accumulation of additional drift in
the integrated signal. Furthermore, the stability analysis for the usual
integral methods are seldom given. Therefore, it is important for the design
of low-drift integral methods. Importantly, the stability and robustness of
integral methods should be analysed.

The algorithms of derivative and integral have been studied by a number of
researchers from different perspectives [3]-[20]. In recent years, the
observer-like differentiators have been developed [15]-[20], and their
stabilities were analysed. Obviously, for the usual observers or
differentiators [15]-[20], it is impossible to estimate the integrals of a
signal. The usual observers or differentiators can estimate the derivatives
of a signal, but not its multiple integrals.

The integral of signal is the infinite sum of differential. The integral of $%
f(x)$ on $[a,b]$ is defined as $I=\int_{a}^{b}f(x)dx=\lim_{\max \{\Delta
x_{i}\rightarrow 0\}}\sum\limits_{{i=1}}^{{n}}f(\xi _{i})\Delta x_{i}$,
where $a=x_{0}<x_{1}<\cdots <x_{n}=b$, $\Delta x_{i}=x_{i}-x_{i-1}$, $\xi
_{i}\in (x_{i-1},x_{i})$, $i=1,\cdots ,n$. Furthermore, The double integral
of $f(x)$ on domain $G$ is defined as $\int_{{}}^{{}}%
\int_{G}^{{}}f(x,y)dxdy=\lim_{\max \{\Delta x_{i}^{2}+\Delta
y_{i}^{2}\rightarrow 0\}}\sum\limits_{{i}}\sum\limits_{{j}}f(\xi _{i},\eta
_{j})\Delta x_{i}\Delta y_{j}$, where $\Delta x_{i}=x_{i}-x_{i-1}$, $\Delta
y_{j}=y_{j}-y_{j-1}$, $\xi _{i}\in (x_{i-1},x_{i})$, $\eta _{j}\in
(y_{j-1},y_{j})$,$i,j=1,\cdots ,n$.

At present, there are no general rules for integrating as there are for
differentiating. There are some numerical methods for estimating signal
integral [12]: i) The trapezoidal rule; ii) Simpson's rule.\emph{\ }For the
above numerical integrating methods, if stochastic noise (especially
non-white noise) exists in signal, and the average value of the noise is not
equal to zero, then such noise leads to the accumulation of additional drift
in the integrated signal. It is not guaranteed that the integrator is
stable. Importantly, it is difficult to analyse the stability and robustness
of these methods. Therefore, the choice of integration scheme has a big
effect on the performance of a system.

The integral operators of $1/s$ and $1/s^{2}$ are irrational and can't be
calculated directly. Researchers presented some approximated methods to
estimate signal integral [13]-[21]. For the integrators design in frequency
domain, an IIR digital integrator was designed by using the Simpson
integration rule and fractional delay filter [3]. In [4], a digital IIR
integrator based on the recursive Romberg integration rule and the
fractional sample delay was investigated. A general theory of the
Newton-Cotes digital integrators by applying the z-transform technique to
the closed-form Newton-Cotes integration formula was presented [5]. In
[7]-[11], several types of digital integrators were designed: non-inverting
integrator, the developed infinite impulse response digital integrators,
low-frequency differential differentiators. However, in all of the
aforementioned studies [3]-[12], the integrators are all linear
approximated, only 1-fold integral can be obtained, and the estimations of
derivative and integral are not considered synchronously. Importantly, there
is no stability analysis. Some integrators were implemented in the hardware,
where the parameters are usually affected by the circumstances, for
instance, the temperature changes in the circuit. Thus, the estimation
precision is affected adversely. Furthermore, they are easily disturbed by
stochastic noise, and the drift phenomenon occurs in such systems. In order
to restrain the stochastic noise, additional filters must be designed.

In [13] and [14], the concept of fractional-order integrator has been
presented, and a rational transfer function is proposed to approximate the
irrational fractional-order integrator $G_{I}\left( s\right) =1/s^{m}$,
where $m$ is a positive real number, and $0<m<1$ is required. The condition
of $0<m<1$ limits the application of the fractional-order integrator.
Usually, the 1-fold and double integrals are required to obtained in many
navigation systems, i.e., the operators $1/s$ and $1/s^{2}$ are required to
be computed. Therefore, this method is not suitable to some navigation
systems.

In this paper, a high-order nonlinear integral-derivative observer is
presented based on finite-time stability [21]-[28] and singular perturbation
technique [29]-[31]. In this derivative-integral observer, a new
distribution of perturbation parameters is arranged. Using this
integral-derivative observer, the multiple integrals and derivatives of a
signal can be obtained synchronously, and the estimation of the time
derivatives and integrals is finite-time stable. The integral-chain
structure increases filter order and improves the stochastic noise
rejection. Importantly, the parameters regulation is only required to be
satisfied with Hurwitz condition. The synchronous dynamical performance of
time derivatives and integrals is helpful for the design of system
controllers (for example, design of proportional-integral-differential (PID)
controller) and the guidance methodologies. Furthermore, for the proposed
integral-derivative observer, the introduction of a differential feedback
term is equivalent that there is a lead compensation (or a dynamic feedback
term) in the integral-derivative observer, and the dynamic performance can
be improved. Therefore, the drift in the integral outputs can be corrected
sufficiently. Importantly, with respect other integrator design, the
stability and robustness in time domain of the presented integral-derivative
observer are analysed. The closed-loop system is stable.

This paper is organized in the following format. In section 2, preliminaries
are introduced. In Section 3, the main results of the presented
integral-derivative observer are presented. In Section 4, the robustness of
the proposed integral-derivative observer is analysed. In Section 5, the
computational analysis and simulations are described. The conclusions are
provided in Section 6.

\bigskip

\textbf{2. Preliminaries and background}

The related concepts are presented here.

\emph{Definition 1 [22]:}\ Let us consider a time-invariant system in the
form of

\begin{equation}
\dot{x}=f\left( x\right) ,f\left( 0\right) =0,x\in R^{n},
\end{equation}%
where $f:D\rightarrow R^{n}$\ is continuous on open neighborhood $D\subseteq
R^{n}$\ of the origin. The origin is said to be a finite-time-stable
equilibrium of the above system if there exists an open neighborhood $%
N\subseteq D$\ of the origin and a function $T_{f}:N\backslash \left\{
0\right\} \rightarrow \left( 0,\infty \right) $, called the settling-time
function, such that the following statements hold: (i)
Finite-time-convergence: For every $x\in N\backslash \left\{ 0\right\} $, $%
\psi ^{x}$\ is the flow starting from $x$ and defined on $\left[
0,T_{f}\left( x\right) \right) $, $\psi ^{x}\left( t\right) \in N\backslash
\left\{ 0\right\} $\ for all $t\in \left[ 0,T_{f}\left( x\right) \right) $,
and $\lim_{t\rightarrow T_{f}\left( x\right) }\psi ^{x}\left( t\right) =0$.

(ii) Lyapunov stability: For every open neighborhood $U_{\varepsilon }$\ of $%
0$\ there exists an open subset $U_{\delta }$\ of $N$\ containing $0$\ such
that, for every $x\in U_{\delta }\backslash \left\{ 0\right\} $, $\psi
^{x}\left( t\right) \in U_{\varepsilon }$\ for all $t\in \left[
0,T_{f}\left( x\right) \right) $.

The origin is said to be a globally finite-time-stable equilibrium if it is
a finite-time-stable equilibrium with $D=N=R^{n}$. Then the system is said
to be finite-time-convergent with respect to the origin.

\emph{Assumption 1:}\ For a system depicted by Equation (1), there exists $%
\rho _{i}\in \left( 0,1\right] ,i=1,\cdots ,n$, and a nonnegative constant $%
\overline{a}$\ such that

\begin{equation}
\left\vert {f}_{j}\left( \widetilde{z}_{1},\widetilde{z}_{2},\cdots ,%
\widetilde{z}_{n}\right) {-f}_{j}\left( \overline{z}_{1},\overline{z}%
_{2},\cdots ,\overline{z}_{n}\right) \right\vert \leq \overline{a}%
\sum\limits_{{i=1}}^{{n}}\left\vert \widetilde{{z}}_{{i}}{-}\overline{z}%
_{i}\right\vert ^{{\rho }_{i}}
\end{equation}%
where $\widetilde{z}_{i},\overline{z}_{i}\in R,i=1,\cdots ,n$, $j=1,\cdots
,n $.

\emph{Remark 1:} There are a number of nonlinear functions capable of
satisfying this assumption. For example, one such function is $x^{\rho _{i}}$%
\ since

\begin{equation*}
\left\vert x^{\rho _{i}}-\overline{x}^{\rho _{i}}\right\vert \leq 2^{1-\rho
_{i}}\left\vert x-\overline{x}\right\vert ^{\rho _{i}},\rho _{i}\in \left(
0,1\right]
\end{equation*}

Moreover, there are smooth functions also satisfying this property. In fact,
it is easy to verify that
\begin{equation*}
\left\vert \sin x-\sin \overline{x}\right\vert \leq 2\left\vert x-\overline{x%
}\right\vert ^{\rho _{i}}
\end{equation*}%
for any $\rho _{i}\in \left( 0,1\right] $.

\emph{Theorem 4.2 [22]:}\ Suppose there exists a continuous function $%
V:R^{n}-R$ such that the following conditions holds:

(i) $V$ is positive definite,

(ii) There exist real numbers $c>0$ and $\theta \in (0,1)$ such that

\begin{equation}
\dot{V}(x)+c(V\left( x\right) )^{\theta }\leq 0
\end{equation}%
Then (1) is globally finite-time stable. Moreover, if $N$ is as in
Definition 1 and $T_{f}$ is the setting time function then

\begin{equation}
T_{f}\left( x\right) \leq \frac{1}{c(1-\theta )}V\left( x\right) ^{1-\theta }
\end{equation}

\emph{Proposition 8.1 [21]:} Let $k_{1},\cdots ,k_{n}>0$ be such that $%
s^{n}+k_{n}s^{n-1}+\cdots +k_{2}s+k_{1}$ is Hurwitz, and consider the system

\begin{eqnarray}
\dot{x}_{i} &=&x_{i+1};i=1,\cdots ,n-1,  \notag \\
\dot{x}_{n} &=&-\sum\limits_{{i=1}}^{{n}}k_{i}\left\vert x_{i}\right\vert
^{\alpha _{i}}sign(x_{i})
\end{eqnarray}%
there exists $\xi \in (0,1)$ such that, for every $\alpha \in (1-\xi ,1)$,
the origin is globally finite-time-stable equilibrium for Equation (5) where
$\alpha _{1},\cdots ,\alpha _{n}$ satisfy

\begin{equation}
\alpha _{i-1}=\frac{\alpha _{i}\alpha _{i+1}}{2\alpha _{i+1}-\alpha _{i}}%
,i=2,\cdots ,n
\end{equation}%
with $\alpha _{n+1}=1$ and $\alpha _{n}=\alpha $.

\emph{Theorem 5.2 [22]:} Consider the perturbed system of (1) following:

\begin{equation}
\overset{\centerdot }{x}=f\left( x\right) +g\left( t,x\left( t\right)
\right) ,x\left( 0\right) =x_{0}
\end{equation}

Suppose there exists a function $V:D\rightarrow R$ such that $V$ is positive
definite and Lipschitz continuous on $D$, and satisfies (3), where $\nu
\subseteq D$ is an open neighborhood of the origin, $c>0$ and $\theta \in (0,%
\frac{1}{2})$. Then there exist $\delta _{0}>0$, $L>0$, $\Gamma >0$, and an
open neighborhood $U$ of origin such that, for every continuous function $%
g:R_{+}\times D\rightarrow R^{n}$ with

\begin{equation}
\delta =\underset{R_{+}\times D}{\sup }\left\Vert g\left( t,x\left( t\right)
\right) \right\Vert \leq \delta _{0}
\end{equation}%
every right maximally defined solution $x$ of (7) with $x(0)\in U$ is
defined on $R_{+}$ and satisfies $x(t)\in U$ for all $t\in R_{+}$ and

\begin{equation}
\left\Vert x\left( t\right) \right\Vert \leq L\delta ^{\gamma },t\geq \Gamma
\end{equation}%
where $\gamma =(1-\theta )/\theta >1$.

\bigskip

\textbf{3. Nonlinear integral-derivative observer}

In the following, finite-time stability [21]-[28] and singular perturbation
technique [29]-[31] will be used to design a high-order nonlinear
integral-derivative observer, and Theorem 1 is presented as follow.

\emph{Theorem 1:} For system

\begin{eqnarray}
\dot{x}_{i} &=&x_{i+1};i=1,\cdots ,n-1  \notag \\
\varepsilon ^{n+1}\dot{x}_{n} &=&-\sum\limits_{{i=1,i\neq p}}^{{n}%
}k_{i}\left\vert \varepsilon ^{i}x_{i}\right\vert ^{\alpha _{i}}sign\left(
x_{i}\right)  \notag \\
&&-k_{p}\left\vert x_{p}-a\left( t\right) \right\vert ^{\alpha
_{p}}sign\left( x_{p}-a\left( t\right) \right)
\end{eqnarray}%
with $p\in \left\{ 2,\cdots ,n\right\} $, if signal $a\left( t\right) $\ is
continuous and ($n-p+1$)th-order derivable, then there exist $\gamma >1$ and
$\Gamma >0$, such that, for $t\geq t_{s}=\varepsilon \Gamma \left( \Xi
(\varepsilon )e\left( {0}\right) \right) $,

\begin{equation}
\left\vert x_{i}-a_{i}\left( t\right) \right\vert \leq L\varepsilon ^{\alpha
_{1}\gamma -i},i=1,\cdots ,n
\end{equation}%
where $a_{p-i}\left( t\right) =\underset{i}{\underbrace{\int_{0}^{t}\cdots
\int_{0}^{\sigma _{2}}}}a\left( \sigma _{1}\right) \underset{i}{\underbrace{%
d\sigma _{1}\cdots d\sigma _{i}}}$, $x_{i}\left( 0\right) =a_{i}\left(
0\right) $, $i=1,\cdots ,p-1$; $x_{p}\left( 0\right) =a_{p}\left( 0\right) $%
; $a_{p}\left( t\right) =a\left( t\right) $; $a_{q}\left( t\right)
=a^{\left( q-p\right) }\left( t\right) $, $q=p+1,\cdots ,n$; $\varepsilon
\in \left( 0,1\right) $ is the perturbation parameter; $L$ is some positive
constant; sup$_{t\in \lbrack 0,\infty )}|a_{i}\left( t\right) |\leq
h_{i}<\infty $, $i=1,\cdots ,n,i\neq p$; sup$_{t\in \lbrack 0,\infty
)}|a^{\left( n-p+1\right) }(t)|\leq {L}_{{a}}<\infty $; $k_{1},\cdots
,k_{p},\cdots ,k_{n}>0$ are selected such that $s^{n}+k_{n}s^{n-1}+\cdots +%
\frac{k_{p}}{\varepsilon ^{p\alpha _{p}}}s^{p-1}+\cdots +k_{2}s+k_{1}$ is
Hurwitz; and $\alpha _{1},\cdots ,\alpha _{n}$ satisfy Equation (6); $\gamma
=(1-\theta )/\theta $, $\theta \in (0,\alpha _{1}/(\alpha _{1}+n+1))$, $%
n\geq 2$; and $\Xi (\varepsilon )=diag\{\varepsilon ,\varepsilon ^{2},\cdots
,\varepsilon ^{n}\}$. $e_{i}=x_{i}-a_{i}\left( t\right) $, $i=1,\cdots ,n$, $%
e=[%
\begin{array}{ccc}
e_{1} & \cdots & e_{n}%
\end{array}%
]^{{T}}$.

\emph{Proof:} The system error between system (10) and the derivatives of $%
a_{1}(t)$ is obtained as follow:

\begin{eqnarray}
e_{i} &=&e_{i+1};i=1,\cdots ,n-1  \notag \\
\varepsilon ^{n+1}\dot{e}_{n} &=&-\sum\limits_{{i=1,i\neq p}}^{{n}%
}k_{i}\left\vert \varepsilon ^{i}e_{i}+\varepsilon ^{i}a_{i}\left( t\right)
\right\vert ^{\alpha _{i}}  \notag \\
&&\times sign\left( e_{i}+a_{i}\left( t\right) \right)  \notag \\
&&-\frac{k_{p}}{\varepsilon ^{p\alpha _{p}}}\left\vert \varepsilon
^{p}e_{p}\right\vert ^{\alpha _{p}}sign\left( e_{p}\right)  \notag \\
&&-\varepsilon ^{n+1}a^{\left( n-p+1\right) }(t)
\end{eqnarray}

Thus, Equation (12) can be rewritten as:

\begin{eqnarray}
\frac{d\varepsilon ^{i}e_{i}}{dt/\varepsilon } &=&\varepsilon
^{i+1}e_{i+1};i=1,\cdots ,n-1  \notag \\
\frac{d\varepsilon ^{n}e_{n}}{dt/\varepsilon } &=&-\sum\limits_{{i=1,i\neq p}%
}^{{n}}k_{i}\left\vert \varepsilon ^{i}e_{i}+\varepsilon ^{i}a_{i}\left(
t\right) \right\vert ^{\alpha _{i}}  \notag \\
&&\times sign\left( e_{i}+a_{i}\left( t\right) \right)  \notag \\
&&-\frac{k_{p}}{\varepsilon ^{p\alpha _{p}}}\left\vert \varepsilon
^{p}e_{p}\right\vert ^{\alpha _{p}}sign\left( e_{p}\right)  \notag \\
&&-\varepsilon ^{n+1}a^{\left( n-p+1\right) }(t)
\end{eqnarray}

Let a coordinate transformation be described as follows:

\begin{eqnarray}
\tau &=&t/\varepsilon ,z_{i}\left( \tau \right) =\varepsilon ^{i}e_{i}\left(
t\right) ,  \notag \\
i &=&1,\cdots ,n,z=[%
\begin{array}{ccc}
z_{1} & \cdots & z_{n}%
\end{array}%
]^{T},  \notag \\
\bar{a}_{i}\left( \tau \right) &=&\varepsilon ^{i}a_{i}\left( t\right)
,i=1,\cdots ,n,i\neq p,  \notag \\
\bar{a}_{n+1}\left( \tau \right) &=&\varepsilon ^{n+1}a^{\left( n-p+1\right)
}(t)
\end{eqnarray}%
Therefore, we obtain $z=\Xi (\varepsilon )e$, and Equation (13) can be
written as

\begin{eqnarray}
\frac{dz_{i}}{d\tau } &=&z_{i+1};i=1,\cdots ,n-1  \notag \\
\frac{dz_{n}}{d\tau } &=&-\sum\limits_{{i=1,i\neq p}}^{{n}}k_{i}\left\vert
z_{i}+\bar{a}_{i}\left( \tau \right) \right\vert ^{\alpha _{i}}sign\left(
z_{i}+\bar{a}_{i}\left( \tau \right) \right)  \notag \\
&&-\frac{k_{p}}{\varepsilon ^{p\alpha _{p}}}\left\vert z_{p}\right\vert
^{\alpha _{p}}sign\left( z_{p}\right) -\bar{a}_{n+1}\left( \tau \right)
\end{eqnarray}

Furthermore, Equation (15) can be rewritten as

\begin{eqnarray}
\frac{dz_{i}}{d\tau } &=&z_{i+1},i=1,\cdots ,n-1  \notag \\
\frac{dz_{n}}{d\tau } &=&-\sum\limits_{{i=1,i\neq p}}^{{n}}k_{i}\left\vert
z_{i}\right\vert ^{\alpha _{i}}sign\left( z_{i}\right)  \notag \\
&&-\frac{k_{p}}{\varepsilon ^{p\alpha _{p}}}\left\vert z_{p}\right\vert
^{\alpha _{p}}sign\left( z_{p}\right)  \notag \\
&&-\sum\limits_{{i=1,i\neq p}}^{{n}}k_{i}\left\{ \left\vert z_{i}+\bar{a}%
_{i}\left( \tau \right) \right\vert ^{\alpha _{i}}sign\left( z_{i}+\bar{a}%
_{i}\left( \tau \right) \right) \right.  \notag \\
&&\left. -\left\vert z_{i}\right\vert ^{\alpha _{i}}sign\left( z_{i}\right)
\right\} -\bar{a}_{n+1}\left( \tau \right)
\end{eqnarray}

Let

\begin{eqnarray}
g_{2}\left( \tau ,z\left( \tau \right) \right) &=&-\sum\limits_{{i=1,i\neq p}%
}^{{n}}k_{i}\left\{ \left\vert z_{i}+\bar{a}_{i}\left( \tau \right)
\right\vert ^{\alpha _{i}}\right.  \notag \\
&&\times sign\left( z_{i}+\bar{a}_{i}\left( \tau \right) \right)  \notag \\
&&\left. -\left\vert z_{i}\right\vert ^{\alpha _{i}}sign\left( z_{i}\right)
\right\} -\bar{a}_{n+1}\left( \tau \right)
\end{eqnarray}

Therefore, from Assumption 1 and Remark 1, we obtain

\begin{eqnarray}
\delta &=&\underset{(\tau ,z)\in R^{n+1}}{\sup }\left\vert g_{2}\left( \tau
,z\left( \tau \right) \right) \right\vert  \notag \\
&\leq &\sum\limits_{{i=1,i\neq p}}^{{n}}2^{1-\alpha _{i}}k_{i}h_{i}^{\alpha
_{i}}\varepsilon ^{i\alpha _{i}}+\varepsilon ^{n+1}L_{a}\leq \varepsilon
^{\rho }\delta _{0}
\end{eqnarray}%
where $\delta _{0}=\sum\limits_{{i=1,i\neq p}}^{{n}}2^{1-\alpha
_{i}}k_{i}h_{i}^{\alpha _{i}}+L_{a}$, and

\begin{equation*}
\rho =\min_{i\in \left\{ 1,\cdots ,n\right\} ,i\neq p}\left\{ \min
\{n+1,i\alpha _{i}\}\right\} =\alpha _{1}
\end{equation*}

In fact, it is checked that the recursive form of Equation (6) may be
rewritten in the non-recursive form

\begin{equation}
\alpha _{i}=\frac{\alpha _{n}}{\left( n-i+1\right) -(n-i)\alpha _{n}}%
,i=1,\cdots ,n
\end{equation}

We will calculate the minimum value of the following expression:

\begin{equation}
i\alpha _{i}=\frac{i\alpha _{n}}{\left( n-i+1\right) -(n-i)\alpha _{n}}%
,i=1,\cdots ,n
\end{equation}

Defining the following function

\begin{equation}
g_{3}(w)=\frac{w\alpha _{n}}{\left( n-w+1\right) -(n-w)\alpha _{n}},w\in
(0,n+1)
\end{equation}%
and taking derivative of $g_{3}(w)$ with respect to variable $w$, we obtain

\begin{equation}
\frac{dg_{3}(w)}{dw}=\frac{\alpha _{n}[\left( n+1\right) -\alpha _{n}n]}{%
[\left( n-w+1\right) -(n-w)\alpha _{n}]^{2}}>0
\end{equation}

Because $\alpha _{n}\in (0,1)$, function $g_{3}(w)$ is monotone increasing.
Moreover, the sequence $\{1,\cdots ,n\}$ is monotone increasing in $(0,n+1)$%
. Therefore,

\begin{equation}
\underset{i\in \left\{ 1,\cdots ,n\right\} ,i\neq p}{\min }\left\{ i\alpha
_{i}\right\} =\alpha _{1}
\end{equation}

Furthermore, because $\varepsilon \in (0,1)$ and $\alpha _{i}\in (0,1)$, $%
i=1,\cdots ,n$, we obtain

\begin{equation}
\underset{i\in \left\{ 1,\cdots ,n\right\} ,i\neq p}{\max }\{\varepsilon
^{i\alpha _{i}}\}=\varepsilon ^{\rho }=\varepsilon ^{\alpha _{1}}
\end{equation}

From Proposition 8.1 in [21], Theorem 5.2 in [22] and Equation (18), there
exist positive constants $\mu $ and $\Gamma \left( z\left( 0\right) \right) $%
, such that

\begin{equation}
\left\Vert z\left( \tau \right) \right\Vert \leq \mu \delta ^{\gamma }\leq
\mu (\varepsilon ^{\alpha _{1}}\delta _{0})^{\gamma },\forall \tau \in
\lbrack \Gamma \left( z\left( 0\right) \right) ,\infty )
\end{equation}

Therefore, from coordinate transformation (14), we obtain

\begin{equation}
\Vert \left[
\begin{array}{ccc}
\varepsilon e_{1} & \cdots & \varepsilon ^{n}e_{n}%
\end{array}%
\right] \Vert \leq \mu (\varepsilon ^{\alpha _{1}}\delta _{0})^{\gamma
},\forall t\in \lbrack \varepsilon \Gamma \left( {\Xi (\varepsilon )e}\left(
{0}\right) \right) ,\infty )
\end{equation}

Thus, the following inequality holds:

\begin{equation}
\left\vert e_{i}\right\vert \leq \mu (\varepsilon ^{\alpha _{1}-\frac{i}{%
\gamma }}\delta _{0})^{\gamma }=L\varepsilon ^{\alpha _{1}\gamma -i},\forall
t\in \lbrack \varepsilon \Gamma \left( {\Xi (\varepsilon )e}\left( {0}%
\right) \right) ,\infty )
\end{equation}%
where $L=\mu \delta _{0}{}^{\gamma }$. To make $\alpha _{1}\gamma -i>1$, $%
i=1,\cdots ,n$, from Theorem 5.2 in [22], we let%
\begin{eqnarray}
\theta &\in &\left( 0,\min \left\{ \alpha _{1}/(\alpha _{1}+n+1),1/2\right\}
\right)  \notag \\
&=&\left( 0,\alpha _{1}/(\alpha _{1}+n+1)\right)
\end{eqnarray}

In fact, from Theorem 4.3 in [26], $\theta $\ can be chosen to be
arbitrarily small. Hence, the requirement that $\theta $\ lies on $\theta
\in \left( 0,\alpha _{1}/(\alpha _{1}+n+1)\right) $\ is not restrictive.
Accordingly, we can obtain $\alpha _{1}[(1-\theta )/\theta ]-n>1$.
Therefore, $\alpha _{1}\gamma -i>1$\ for $i=1,\cdots ,n$. The choice of $%
\theta $ leads to $\alpha _{1}\gamma -i>1$ in (27) which implies that for $%
\varepsilon <1$, the ultimate bound (27) on the estimation error is of
higher order than the perturbation. This concludes the proof. $\blacksquare $

\bigskip

In order to guarantee the stability of integral derivative observer (10),
polynomial

\begin{equation*}
s^{n}+k_{n}s^{n-1}+\cdots +\frac{k_{p}}{\varepsilon ^{p\alpha _{p}}}%
s^{p-1}+\cdots +k_{2}s+k_{1}
\end{equation*}%
must be Hurwitz for $\varepsilon \in \left( 0,1\right) $ and the bounded
constants $k_{i}>0$, $i=1,\cdots ,n$. In the following, we will discuss the
Routh-Hurwitz Stability Criterion of the polynomial. It will be found that $%
s^{n}+k_{n}s^{n-1}+\cdots +\frac{k_{p}}{\varepsilon ^{p\alpha _{p}}}%
s^{p-1}+\cdots +k_{2}s+k_{1}$ can't be Hurwitz for arbitrary integers $n$
and $p$ with $\varepsilon \in \left( 0,1\right) $. Fortunately, for some
integers $n$ and $p$, Theorem 1 still holds, and it is satisfied with almost
all engineering applications. For instance, the position and velocity can be
estimated from the acceleration signal, or position and acceleration can be
obtained from the velocity signal.

\bigskip

\emph{Lemma 1: }The following selections of $n$ and $p$ can make polynomial $%
s^{n}+k_{n}s^{n-1}+\cdots +\frac{k_{p}}{\varepsilon ^{p\alpha _{p}}}%
s^{p-1}+\cdots +k_{2}s+k_{1}$ Hurwitz, where $\varepsilon \in \left(
0,1\right) $ and the bounded constants $k_{i}>0$, $i=1,\cdots ,n$:

a) $n=2$ and $p=2$;

b) $n=3$ and $p\in \left\{ 2,3\right\} $;

c) $n=4$ and $p=3$.

\emph{Proof:} In the following, for polynomial $s^{n}+k_{n}s^{n-1}+\cdots +%
\frac{k_{p}}{\varepsilon ^{p\alpha _{p}}}s^{p-1}+\cdots +k_{2}s+k_{1}$, we
will search for integers $n$ and $p$ to satisfy the Routh-Hurwitz Stability
Criterion. It is known that Routh table is the nested structure. Therefore,
if there is an integer $N$, when $n=N$, for all $p\in \left\{ 2,\cdots
,N\right\} $ and $\varepsilon \in \left( 0,1\right) $, such that $%
s^{n}+k_{n}s^{n-1}+\cdots +\frac{k_{p}}{\varepsilon ^{p\alpha _{p}}}%
s^{p-1}+\cdots +k_{2}s+k_{1}$ is not Hurwitz, then when $n\geq N$, this
statement still holds.

\emph{1)} When $n=2$ , for polynomial $s^{2}+\bar{k}_{2}s+k_{1}$, the Routh
table is

\begin{equation}
\begin{array}{ccc}
s^{2} & 1 & k_{1} \\
s^{1} & \bar{k}_{2} & 0 \\
s^{0} & k_{1} & 0%
\end{array}%
\end{equation}

Obviously, polynomial $s^{2}+\bar{k}_{2}s+k_{1}$ is Hurwitz if $k_{1}>0,\bar{%
k}_{2}>0$. Therefore, when $n=2$, $p=2$ and $\bar{k}_{2}=\frac{k_{2}}{%
\varepsilon ^{2\alpha _{2}}}$, polynomial $s^{2}+\frac{k_{2}}{\varepsilon
^{2\alpha _{2}}}s+k_{1}$ can be Hurwitz for arbitrary $\varepsilon \in
\left( 0,1\right) $.

\emph{2)} When $n=3$, for polynomial $s^{3}+\bar{k}_{3}s^{2}+\bar{k}%
_{2}s+k_{1}$, the Routh table is

\begin{equation}
\begin{array}{ccc}
s^{3} & 1 & \bar{k}_{2} \\
s^{2} & \bar{k}_{3} & k_{1} \\
s^{1} & \frac{\bar{k}_{3}\bar{k}_{2}-k_{1}}{\bar{k}_{3}} & 0 \\
s^{0} & k_{1} & 0%
\end{array}%
\end{equation}

Polynomial $s^{3}+\bar{k}_{3}s^{2}+\bar{k}_{2}s+k_{1}$ is Hurwitz if $%
k_{1}>0,\bar{k}_{3}>0$ and $\bar{k}_{2}\bar{k}_{3}>k_{1}$. Therefore, when $%
n=3$, $p=3$, $\bar{k}_{2}=k_{2}$ and $\bar{k}_{3}=\frac{k_{3}}{\varepsilon
^{3\alpha _{3}}}$, polynomial $s^{3}+\frac{k_{3}}{\varepsilon ^{3\alpha _{3}}%
}s^{2}+k_{2}s+k_{1}$ can be Hurwitz for arbitrary $\varepsilon \in \left(
0,1\right) $; when $n=3$, $p=2$, $\bar{k}_{2}=\frac{k_{2}}{\varepsilon
^{2\alpha _{2}}}$ and $\bar{k}_{3}=k_{3}$, polynomial $s^{3}+k_{3}s^{2}+%
\frac{k_{2}}{\varepsilon ^{2\alpha _{2}}}s+k_{1}$ can be Hurwitz for
arbitrary $\varepsilon \in \left( 0,1\right) $.

\emph{3)} When $n=4$, for polynomial $s^{4}+\bar{k}_{4}s^{3}+\bar{k}%
_{3}s^{2}+\bar{k}_{2}s+k_{1}$, the Routh table is

\begin{equation}
\begin{array}{cccc}
s^{4} & 1 & \bar{k}_{3} & k_{1} \\
s^{3} & \bar{k}_{4} & \bar{k}_{2} & 0 \\
s^{2} & A_{1} & k_{1} & 0 \\
s^{1} & B_{1} & 0 & 0 \\
s^{0} & k_{1} & 0 & 0%
\end{array}%
\end{equation}%
where

\begin{equation}
A_{1}=\frac{\bar{k}_{4}\bar{k}_{3}-\bar{k}_{2}}{\bar{k}_{4}},B_{1}=\frac{%
\frac{\bar{k}_{4}\bar{k}_{3}-\bar{k}_{2}}{\bar{k}_{4}}\bar{k}_{2}-\bar{k}%
_{4}k_{1}}{\frac{\bar{k}_{4}\bar{k}_{3}-\bar{k}_{2}}{\bar{k}_{4}}}
\end{equation}

Polynomial $s^{4}+\bar{k}_{4}s^{3}+\bar{k}_{3}s^{2}+\bar{k}_{2}s+k_{1}$ is
Hurwitz if $k_{1}>0,\bar{k}_{4}>0,\bar{k}_{4}\bar{k}_{3}>\bar{k}_{2}$ and $%
\bar{k}_{4}\bar{k}_{3}\bar{k}_{2}>\bar{k}_{4}^{2}k_{1}+\bar{k}_{2}^{2}$.
Therefore, only when $n=4$, $p=3$, $\bar{k}_{2}=k_{2}$, $\bar{k}_{3}=\frac{%
k_{3}}{\varepsilon ^{3\alpha _{3}}}$ and $\bar{k}_{4}=k_{4}$, polynomial $%
s^{4}+k_{4}s^{3}+\frac{k_{3}}{\varepsilon ^{3\alpha _{3}}}s^{2}+k_{2}s+k_{1}$
can be Hurwitz for arbitrary $\varepsilon \in \left( 0,1\right) $.

\emph{4)} When $n=5$, for polynomial $s^{5}+\bar{k}_{5}s^{4}+\bar{k}%
_{4}s^{3}+\bar{k}_{3}s^{2}+\bar{k}_{2}s+k_{1}$, the Routh table is

\begin{equation}
\begin{array}{cccc}
s^{5} & 1 & \bar{k}_{4} & \bar{k}_{2} \\
s^{4} & \bar{k}_{5} & \bar{k}_{3} & k_{1} \\
s^{3} & A_{1} & A_{2} & 0 \\
s^{2} & B_{1} & k_{1} & 0 \\
s^{1} & C_{1} & 0 & 0 \\
s^{0} & k_{1} & 0 & 0%
\end{array}%
\end{equation}%
where

\begin{eqnarray}
A_{1} &=&\frac{\bar{k}_{5}\bar{k}_{4}-\bar{k}_{3}}{\bar{k}_{5}},A_{2}=\frac{%
\bar{k}_{5}\bar{k}_{2}-k_{1}}{\bar{k}_{5}},  \notag \\
B_{1} &=&\frac{\frac{\bar{k}_{5}\bar{k}_{4}-\bar{k}_{3}}{\bar{k}_{5}}\bar{k}%
_{3}-\left( \bar{k}_{5}\bar{k}_{2}-k_{1}\right) }{\frac{\bar{k}_{5}\bar{k}%
_{4}-\bar{k}_{3}}{\bar{k}_{5}}},  \notag \\
C_{1} &=&\frac{\frac{\frac{\bar{k}_{5}\bar{k}_{4}-\bar{k}_{3}}{\bar{k}_{5}}%
\bar{k}_{3}-\left( \bar{k}_{5}\bar{k}_{2}-k_{1}\right) }{\frac{\bar{k}_{5}%
\bar{k}_{4}-\bar{k}_{3}}{\bar{k}_{5}}}\frac{\bar{k}_{5}\bar{k}_{2}-k_{1}}{%
\bar{k}_{5}}-\frac{\bar{k}_{5}\bar{k}_{4}-\bar{k}_{3}}{\bar{k}_{5}}k_{1}}{%
B_{1}}
\end{eqnarray}

Polynomial $s^{5}+\bar{k}_{5}s^{4}+\bar{k}_{4}s^{3}+\bar{k}_{3}s^{2}+\bar{k}%
_{2}s+k_{1}$ is Hurwitz if $k_{1}>0,\bar{k}_{5}>0,\bar{k}_{5}\bar{k}_{4}>%
\bar{k}_{3},\bar{k}_{5}\bar{k}_{2}>k_{1},\bar{k}_{4}\bar{k}_{3}+k_{1}>\bar{k}%
_{5}\bar{k}_{2}+\frac{\bar{k}_{3}^{2}}{k_{5}}$ and $\bar{k}_{3}>\frac{\bar{k}%
_{5}\bar{k}_{4}-\bar{k}_{3}}{\bar{k}_{5}\bar{k}_{2}-k_{1}}k_{1}+\frac{\bar{k}%
_{5}\bar{k}_{2}-k_{1}}{\bar{k}_{5}\bar{k}_{4}-\bar{k}_{3}}\bar{k}_{5}$. For
arbitrary large $\bar{k}_{p}=\frac{k_{p}}{\varepsilon ^{p\alpha _{p}}}$ and
all $p\in \left\{ 2,3,4,5\right\} $, polynomial $s^{5}+\cdots +\bar{k}%
_{p}s^{p-1}+\cdots +k_{1}$ can't be Hurwitz. Therefore, for $n=5$, all $p\in
\left\{ 2,3,4,5\right\} $ and arbitrary $\varepsilon \in \left( 0,1\right) $%
, polynomial $s^{5}+\cdots +\frac{k_{p}}{\varepsilon ^{p\alpha _{p}}}%
s^{p-1}+\cdots +k_{1}$ can't be Hurwitz.

Accordingly, we find an integer $N=5$, when $n=5$, for all $p\in \left\{
2,3,4,5\right\} $ and arbitrary large $\frac{k_{p}}{\varepsilon ^{p\alpha
_{p}}}$, polynomial $s^{5}+\cdots +\frac{k_{p}}{\varepsilon ^{p\alpha _{p}}}%
s^{p-1}+\cdots +k_{1}$ can't be Hurwitz. It is known that Routh table is the
nested structure. Therefore, when $n\geq 5$, for all $p\in \left\{ 2,\cdots
,n\right\} $ and arbitrary $\varepsilon \in \left( 0,1\right) $, polynomial $%
s^{n}+k_{n}s^{n-1}+\cdots +\frac{k_{p}}{\varepsilon ^{p\alpha _{p}}}%
s^{p-1}+\cdots +k_{2}s+k_{1}$ can't be Hurwitz. $\blacksquare $

\bigskip

\emph{Remark 2: Perturbation terms }$\varepsilon $\emph{\ and }$\delta $

From the analysis above, for $\varepsilon \in \left( 0,1\right) $ and the
bounded constants $k_{i}>0$, $i=1,\cdots ,n$, there exist $n$ and $p$, such
that polynomial $s^{n}+k_{n}s^{n-1}+\cdots +\frac{k_{p}}{\varepsilon
^{p\alpha _{p}}}s^{p-1}+\cdots +k_{2}s+k_{1}$ is Hurwitz. For the following
perturbation term in Equation (18)

\begin{equation}
\delta =\underset{(\tau ,z)\in R^{n+1}}{\sup }\left\vert g_{2}\left( \tau
,z\left( \tau \right) \right) \right\vert \leq \varepsilon ^{\rho }\delta
_{0}
\end{equation}%
where $\delta _{0}=\sum\limits_{{i=1,i\neq p}}^{{n}}2^{1-\alpha
_{i}}k_{i}h_{i}^{\alpha _{i}}+L_{a}$, and

\begin{equation*}
\rho =\min_{i\in \left\{ 1,\cdots ,n\right\} ,i\neq p}\left\{ i\alpha
_{i}\right\} =a_{1}
\end{equation*}%
the term $\delta _{0}$ is bounded. Furthermore, $\underset{\varepsilon
\rightarrow 0}{\lim }\delta =0$.

\emph{Remark 3: }The following system can also implement the estimation of
the time derivatives and integrals:

\begin{eqnarray}
\dot{x}_{i} &=&x_{i+1},i=1,\cdots ,n-1  \notag \\
\varepsilon ^{n+1}\dot{x}_{n} &=&-\sum\limits_{{i=1,i\neq p}}^{{n}%
}k_{i}\left\vert \varepsilon ^{i}x_{i}\right\vert ^{\alpha _{i}}sign\left(
x_{i}\right)  \notag \\
&&-k_{p}\left\vert \varepsilon ^{p}\left( x_{p}-a\left( t\right) \right)
\right\vert ^{\alpha _{p}}sign\left( x_{p}-a\left( t\right) \right)
\end{eqnarray}%
where, $k_{1},\cdots ,k_{n}>0$ are selected such that

\begin{equation*}
s^{n}+k_{n}s^{n-1}+\cdots +k_{2}s+k_{1}
\end{equation*}%
is Hurwitz. However, the error term $\varepsilon ^{p}\left( x_{p}-a\left(
t\right) \right) $ in Equation (36) is very small, and its corresponding
error feedback is quite weak. Therefore, the system convergence is slow. In
integral-derivative observer (10), the gain $k_{p}$ of the error term $%
x_{p}-a\left( t\right) $ is suitable, thus, the error feedback is more
efficient.$\blacksquare $

\bigskip

From Theorem 1 and Lemma 1, we can obtain the exact forms of
integral-derivative observers, and a Corollary is presented as follow.

\bigskip

\emph{Corollary 1: }There exist the following four types of nonlinear
integral-derivative observers:

\emph{i. 1-fold-integral observer}

When $n=2$ and $p=2$, Equation (10) can be written as

\begin{eqnarray}
\dot{x}_{1} &=&x_{2}  \notag \\
\varepsilon ^{3}\dot{x}_{2} &=&-k_{1}\left\vert \varepsilon x_{1}\right\vert
^{\alpha _{1}}sign\left( x_{1}\right)  \notag \\
&&-k_{2}\left\vert x_{2}-a\left( t\right) \right\vert ^{\alpha
_{2}}sign\left( x_{2}-a\left( t\right) \right)
\end{eqnarray}%
with the conclusion that, for $t\geq t_{s}$,

\begin{equation}
\left\vert x_{1}-a_{1}\left( t\right) \right\vert \leq L\varepsilon ^{\alpha
_{1}\gamma -1},\left\vert x_{2}-a_{2}\left( t\right) \right\vert \leq
L\varepsilon ^{\alpha _{1}\gamma -2}
\end{equation}%
where $a_{1}\left( t\right) =\int_{0}^{t}a\left( \sigma \right) d\sigma $, $%
a_{2}\left( t\right) =a\left( t\right) $; $x_{1}\left( 0\right) =a_{1}\left(
0\right) $, $x_{2}\left( 0\right) =a_{2}\left( 0\right) $; $\varepsilon \in
\left( 0,1\right) $ is the perturbation parameter; $\left\vert \dot{a}%
(t)\right\vert \leq L_{a}$, $L_{a}$ is a positive constant; $k_{1}>0,k_{2}>0$%
; $\alpha _{1}=\frac{\alpha _{2}}{2-\alpha _{2}},\alpha _{2}\in \left(
0,1\right) $; $L$ is some positive constant; $\gamma =(1-\theta )/\theta $, $%
\theta \in (0,\alpha _{1}/(\alpha _{1}+3))$. It is a 1-fold integral
observer, which can obtain the 1-fold integral of signal $a\left( t\right) $.

\bigskip

\emph{ii. First-order-derivative 1-fold-integral observer}

When $n=3$ and $p=2$, Equation (10) can be written as

\begin{eqnarray}
\dot{x}_{1} &=&x_{2}  \notag \\
\dot{x}_{2} &=&x_{3}  \notag \\
\varepsilon ^{4}\dot{x}_{3} &=&-k_{1}\left\vert \varepsilon x_{1}\right\vert
^{\alpha _{1}}sign\left( x_{1}\right)  \notag \\
&&-k_{2}\left\vert x_{2}-a\left( t\right) \right\vert ^{\alpha
_{2}}sign\left( x_{2}-a\left( t\right) \right)  \notag \\
&&-k_{3}\left\vert \varepsilon ^{3}x_{3}\right\vert ^{\alpha _{3}}sign\left(
x_{3}\right)
\end{eqnarray}%
with the conclusion that, for $t\geq t_{s}$,

\begin{equation}
\left\vert x_{i}-a_{i}\left( t\right) \right\vert \leq L\varepsilon ^{\alpha
_{1}\gamma -i},i=1,2,3
\end{equation}%
where $a_{1}\left( t\right) =\int_{0}^{t}a\left( \sigma \right) d\sigma $, $%
a_{2}\left( t\right) =a\left( t\right) $, $a_{3}\left( t\right) =\dot{a}%
\left( t\right) $; $x_{1}\left( 0\right) =a_{1}\left( 0\right) $, $%
x_{2}\left( 0\right) =a_{2}\left( 0\right) $; $\varepsilon \in \left(
0,1\right) $ is the perturbation parameter; $k_{1}>0,k_{3}>0$ and $%
k_{2}>\varepsilon ^{2\alpha _{2}}\frac{k_{1}}{k_{3}};$ $\alpha _{1}=\frac{%
\alpha _{2}\alpha _{3}}{2\alpha _{3}-\alpha _{2}},\alpha _{2}=\frac{\alpha
_{3}}{2-\alpha _{3}},\alpha _{3}\in \left( 0,1\right) $; $L$ is some
positive constant; and $\gamma =(1-\theta )/\theta $, $\theta \in (0,\alpha
_{1}/(\alpha _{1}+4))$. It is an integral-derivative observer, which can
obtain the derivative and integral, respectively, of signal $a\left(
t\right) $.

\bigskip

\emph{iii. Double integral observer}

When $n=3$ and $p=3$, Equation (10) can be written as

\begin{eqnarray}
\dot{x}_{1} &=&x_{2}  \notag \\
\dot{x}_{2} &=&x_{3}  \notag \\
\varepsilon ^{4}\dot{x}_{3} &=&-k_{1}\left\vert \varepsilon x_{1}\right\vert
^{\alpha _{1}}sign\left( x_{1}\right) -k_{2}\left\vert \varepsilon
^{2}x_{2}\right\vert ^{\alpha _{2}}sign\left( x_{2}\right)  \notag \\
&&-k_{3}\left\vert x_{3}-a\left( t\right) \right\vert ^{\alpha
_{3}}sign\left( x_{3}-a\left( t\right) \right)
\end{eqnarray}%
with the conclusion that, for $t\geq t_{s}$,

\begin{equation}
\left\vert x_{i}-a_{i}\left( t\right) \right\vert \leq L\varepsilon ^{\alpha
_{1}\gamma -i},i=1,2,3
\end{equation}%
where $a_{1}\left( t\right) =\int_{0}^{t}\int_{0}^{\sigma _{2}}a\left(
\sigma _{1}\right) d\sigma _{1}d\sigma _{2}$, $a_{2}\left( t\right)
=\int_{0}^{t}a\left( \sigma _{1}\right) d\sigma _{1}$, $a_{3}\left( t\right)
=a\left( t\right) $; $x_{1}\left( 0\right) =a_{1}\left( 0\right) $, $%
x_{2}\left( 0\right) =a_{2}\left( 0\right) $, $x_{3}\left( 0\right)
=a_{3}\left( 0\right) $; $\varepsilon \in \left( 0,1\right) $ is the
perturbation parameter; $k_{1}>0,k_{3}>0$ and $k_{2}>\varepsilon ^{3\alpha
_{3}}\frac{k_{1}}{k_{3}}$; $\alpha _{1}=\frac{\alpha _{2}\alpha _{3}}{%
2\alpha _{3}-\alpha _{2}}$, $\alpha _{2}=\frac{\alpha _{3}}{2-\alpha _{3}}$,
$\alpha _{3}\in \left( 0,1\right) $; $L$ is some positive constant; $\gamma
=(1-\theta )/\theta $, $\theta \in (0,\alpha _{1}/(\alpha _{1}+4))$. It is a
double integral observer, which can obtain the 1-fold and double integrals,
respectively, of signal $a\left( t\right) $.

\bigskip

\emph{iv. First--order-derivative double-integral observer}

When $n=4$ and $p=3$, Equation (10) can be written as

\begin{eqnarray}
\dot{x}_{1} &=&x_{2}  \notag \\
\dot{x}_{2} &=&x_{3}  \notag \\
\dot{x}_{3} &=&x_{4}  \notag \\
\varepsilon ^{5}\dot{x}_{4} &=&-k_{1}\left\vert \varepsilon x_{1}\right\vert
^{\alpha _{1}}sign\left( x_{1}\right) -k_{2}\left\vert \varepsilon
^{2}x_{2}\right\vert ^{\alpha _{2}}sign\left( x_{2}\right)  \notag \\
&&-k_{3}\left\vert x_{3}-a\left( t\right) \right\vert ^{\alpha
_{3}}sign\left( x_{3}-a\left( t\right) \right)  \notag \\
&&-k_{4}\left\vert \varepsilon ^{4}x_{4}\right\vert ^{\alpha _{4}}sign\left(
x_{4}\right)
\end{eqnarray}%
with the conclusion that, for $t\geq t_{s}$,

\begin{equation}
\left\vert x_{i}-a_{i}\left( t\right) \right\vert \leq L\varepsilon ^{\alpha
_{1}\gamma -i},i=1,2,3,4
\end{equation}%
where $a_{1}\left( t\right) =\int_{0}^{t}\int_{0}^{\sigma _{2}}a\left(
\sigma _{1}\right) d\sigma _{1}d\sigma _{2},a_{2}\left( t\right)
=\int_{0}^{t}a\left( \sigma _{1}\right) d\sigma _{1},a_{3}\left( t\right)
=a\left( t\right) ,a_{4}\left( t\right) =\dot{a}\left( t\right) $; $%
x_{1}\left( 0\right) =a_{1}\left( 0\right) $, $x_{2}\left( 0\right)
=a_{2}\left( 0\right) $, $x_{3}\left( 0\right) =a_{3}\left( 0\right) $; $%
\varepsilon \in \left( 0,1\right) $ is the perturbation parameter; $%
k_{1}>0,k_{4}>0,k_{3}>\varepsilon ^{3\alpha _{3}}\frac{k_{2}}{k_{4}}$ and $%
k_{2}>\varepsilon ^{3\alpha _{3}}\frac{k_{4}^{2}k_{1}+k_{2}^{2}}{k_{4}k_{3}}$%
; $\alpha _{1}=\frac{\alpha _{2}\alpha _{3}}{2\alpha _{3}-\alpha _{2}}$, $%
\alpha _{2}=\frac{\alpha _{3}\alpha _{4}}{2\alpha _{4}-\alpha _{3}}$, $%
\alpha _{3}=\frac{\alpha _{4}}{2-\alpha _{4}}$, $\alpha _{4}\in \left(
0,1\right) $; $L$ is some positive constant; $\gamma =(1-\theta )/\theta $, $%
\theta \in (0,\alpha _{1}/(\alpha _{1}+5))$. It is an integral-derivative
observer, which can obtain the 1-fold, double integrals and first-order
derivative, respectively, of signal $a\left( t\right) $.

\bigskip

\textbf{4. Robustness analysis of nonlinear time integral-derivative observer%
}

In a realistic problem, signal $a(t)$ in integral-derivative observer system
(10) might represent an ideal signal without any disturbance, while
stochastic disturbances exist in almost all signals. The following theorem
concerns the robustness behavior of the presented integral-observer under
bounded perturbations.

\emph{Theorem 2:} For integral-derivative observer (10), if the disturbance
exists in signal $a\left( t\right) $, i.e., $a\left( t\right) =a_{0}\left(
t\right) +d\left( t\right) $, where $a_{0}\left( t\right) $ is the desired
signal, $d\left( t\right) $ is the bounded stochastic disturbance, and sup$%
_{t\in \lbrack 0,\infty )}\left\vert d\left( t\right) \right\vert \leq
L_{d}<\infty $, then there exist $\gamma >1$ and $\Gamma >0$, such that, for
$t\geq \varepsilon \Gamma \left( \Xi (\varepsilon )e\left( {0}\right)
\right) $,

\begin{equation}
\left\vert x_{i}-a_{0i}\left( t\right) \right\vert \leq L(\delta
_{di})^{\gamma },i=1,\cdots ,n
\end{equation}%
where

\begin{equation*}
a_{0(p-i)}\left( t\right) =\underset{i}{\underbrace{\int_{0}^{t}\cdots
\int_{0}^{\sigma _{2}}}}a_{0}\left( \sigma _{1}\right) \underset{i}{%
\underbrace{d\sigma _{1}\cdots d\sigma _{i}}}
\end{equation*}%
$x_{i}\left( 0\right) =a_{0i}\left( 0\right) $, $i=1,\cdots ,p-1$; $%
a_{0p}\left( t\right) =a_{0}\left( t\right) $; $a_{0q}\left( t\right)
=a_{0}^{\left( q-p\right) }\left( t\right) $, $q=p+1,\cdots ,n$; $L$ is some
positive constant; $\delta _{di}=\varepsilon ^{\alpha _{1}-\frac{i}{\gamma }%
}+\frac{2^{1-\alpha _{p}}}{\delta _{0}}k_{p}L_{d}^{\alpha _{p}}\varepsilon
^{-\frac{i}{\gamma }}$, and $\delta _{di}\in (0,1)$, $i=1,\cdots ,n$; $%
\varepsilon \in \left( 0,1\right) $, and $L_{d}<\left( \frac{1-\varepsilon
^{\alpha _{1}}}{2^{1-\alpha _{p}}k_{p}}\delta _{0}\right) ^{\frac{1}{\alpha
_{p}}}$; $\delta _{0}=\sum\limits_{{i=1,i\neq p}}^{{n}}2^{1-\alpha
_{i}}k_{i}h_{i}^{\alpha _{i}}+L_{a}$; sup$_{t\in \lbrack 0,\infty
)}|a_{0i}\left( t\right) |\leq h_{i}<\infty $, $i=1,\cdots ,n$, $i\neq p$;
sup$_{t\in \lbrack 0,\infty )}|a_{0}^{\left( n-p+1\right) }(t)|\leq {L}_{{a}%
}<\infty $; $\gamma =(1-\theta )/\theta $,

\begin{equation*}
\theta \in \left( 0,\min \left\{ \frac{1}{\frac{(n+1)\log \varepsilon }{\log
(\varepsilon ^{\alpha _{1}}+\frac{2^{1-\alpha _{p}}}{\delta _{0}}%
k_{p}L_{d}^{\alpha _{p}})}+1},\frac{1}{2}\right\} \right) ,n\geq 2;
\end{equation*}%
$\Xi (\varepsilon )=diag\{\varepsilon ,\varepsilon ^{2},\cdots ,\varepsilon
^{n}\}$, and $e_{i}=x_{i}-a_{0i}\left( t\right) $, $i=1,\cdots ,n$, $e=[%
\begin{array}{ccc}
e_{1} & \cdots & e_{n}%
\end{array}%
]^{{T}}$.

\emph{Proof:} The system error between system (10) and the derivatives of $%
a_{01}(t)$ is given by:

\begin{eqnarray}
e_{i} &=&e_{i+1};i=1,\cdots ,n-1  \notag \\
\varepsilon ^{n+1}\dot{e}_{n} &=&-\sum\limits_{{i=1,i\neq p}}^{{n}%
}k_{i}\left\vert \varepsilon ^{i}e_{i}+\varepsilon ^{i}a_{0i}\left( t\right)
\right\vert ^{\alpha _{i}}  \notag \\
&&\times sign\left( e_{i}+a_{0i}\left( t\right) \right)  \notag \\
&&-\frac{k_{p}}{\varepsilon ^{p\alpha _{p}}}\left\vert \varepsilon
^{p}e_{p}-\varepsilon ^{p}d\left( t\right) \right\vert ^{\alpha _{p}}  \notag
\\
&&\times sign\left( e_{p}-d\left( t\right) \right) -\varepsilon
^{n+1}a_{0}^{\left( n-p+1\right) }(t)
\end{eqnarray}

The Equation (46) can be rewritten as:

\begin{eqnarray}
\frac{d\varepsilon ^{i}e_{i}}{dt/\varepsilon } &=&\varepsilon
^{i+1}e_{i+1};i=1,\cdots ,n-1  \notag \\
\frac{d\varepsilon ^{n}e_{n}}{dt/\varepsilon } &=&-\sum\limits_{{i=1,i\neq p}%
}^{{n}}k_{i}\left\vert \varepsilon ^{i}e_{i}+\varepsilon ^{i}a_{0i}\left(
t\right) \right\vert ^{\alpha _{i}}  \notag \\
&&\times sign\left( e_{i}+a_{0i}\left( t\right) \right)  \notag \\
&&-\frac{k_{p}}{\varepsilon ^{p\alpha _{p}}}\left\vert \varepsilon
^{p}e_{p}-\varepsilon ^{p}d\left( t\right) \right\vert ^{\alpha _{p}}  \notag
\\
&&\times sign\left( e_{p}-d\left( t\right) \right) -\varepsilon
^{n+1}a_{0}^{\left( n-p+1\right) }(t)
\end{eqnarray}

Let%
\begin{eqnarray}
\tau &=&t/\varepsilon ,z_{i}\left( \tau \right) =\varepsilon ^{i}e_{i}\left(
t\right) ,\bar{a}_{i}\left( \tau \right) =\varepsilon ^{i}a_{0i}\left(
t\right) ,  \notag \\
i &=&1,\cdots ,n,z=[%
\begin{array}{ccc}
z_{1} & \cdots & z_{n}%
\end{array}%
]^{T},  \notag \\
\bar{a}_{n+1}\left( \tau \right) &=&\varepsilon ^{n+1}a_{0}^{\left(
n-p+1\right) }(t),\bar{d}\left( \tau \right) =\varepsilon ^{p}d\left(
t\right)
\end{eqnarray}%
therefore, we have $z=\Xi (\varepsilon )e$. The Equation (47) can be written
as%
\begin{eqnarray}
\frac{dz_{i}}{d\tau } &=&z_{i+1};i=1,\cdots ,n-1  \notag \\
\frac{dz_{n}}{d\tau } &=&\sum\limits_{{i=1,i\neq p}}^{{n}}k_{i}\left\vert
z_{i}+\bar{a}_{i}\left( \tau \right) \right\vert ^{\alpha _{i}}sign\left(
z_{i}+\bar{a}_{i}\left( \tau \right) \right)  \notag \\
&&-\frac{k_{p}}{\varepsilon ^{p\alpha _{p}}}\left\vert z_{p}-\bar{d}\left(
\tau \right) \right\vert ^{\alpha _{p}}sign\left( z_{p}-\bar{d}\left( \tau
\right) \right)  \notag \\
&&-\bar{a}_{n+1}\left( \tau \right)
\end{eqnarray}

Furthermore, Equation (49) can be rewritten as

\begin{eqnarray}
\frac{dz_{i}}{d\tau } &=&z_{i+1},i=1,\cdots ,n-1  \notag \\
\frac{dz_{n}}{d\tau } &=&-\sum\limits_{{i=1,i\neq p}}^{{n}}k_{i}\left\vert
z_{i}\right\vert ^{\alpha _{i}}sign\left( z_{i}\right)  \notag \\
&&-\frac{k_{p}}{\varepsilon ^{p\alpha _{p}}}\left\vert z_{p}\right\vert
^{\alpha _{p}}sign\left( z_{p}\right)  \notag \\
&&-\frac{k_{p}}{\varepsilon ^{p\alpha _{p}}}\left\{ \left\vert z_{p}-\bar{d}%
\left( \tau \right) \right\vert ^{\alpha _{p}}sign\left( z_{p}-\bar{d}\left(
\tau \right) \right) \right.  \notag \\
&&\left. -\left\vert z_{p}\right\vert ^{\alpha _{p}}sign\left( z_{p}\right)
\right\}  \notag \\
&&-\sum\limits_{{i=1,i\neq p}}^{{n}}k_{i}\left\{ \left\vert z_{i}+\bar{a}%
_{i}\left( \tau \right) \right\vert ^{\alpha _{i}}sign\left( z_{i}+\bar{a}%
_{i}\left( \tau \right) \right) \right.  \notag \\
&&\left. -\left\vert z_{i}\right\vert ^{\alpha _{i}}sign\left( z_{i}\right)
\right\} -\bar{a}_{n+1}\left( \tau \right)
\end{eqnarray}

Let

\begin{eqnarray}
g_{2}\left( \tau ,z\left( \tau \right) \right) &=&-\sum\limits_{{i=1,i\neq p}%
}^{{n}}k_{i}\left\{ \left\vert z_{i}+\bar{a}_{i}\left( \tau \right)
\right\vert ^{\alpha _{i}}\right.  \notag \\
&&\times sign\left( z_{i}+\bar{a}_{i}\left( \tau \right) \right)  \notag \\
&&\left. -\left\vert z_{i}\right\vert ^{\alpha _{i}}sign\left( z_{i}\right)
\right\} -\bar{a}_{n+1}\left( \tau \right)  \notag \\
&&-\frac{k_{p}}{\varepsilon ^{p\alpha _{p}}}\left\{ \left\vert z_{p}-\bar{d}%
\left( \tau \right) \right\vert ^{\alpha _{p}}\right.  \notag \\
&&\left. \times sign\left( z_{p}-\bar{d}\left( \tau \right) \right)
-\left\vert z_{p}\right\vert ^{\alpha _{p}}sign\left( z_{p}\right) \right\}
\end{eqnarray}

Therefore, from Assumption 1 and Remark 1, we obtain

\begin{eqnarray}
\delta &=&\underset{(\tau ,z)\in R^{n+1}}{\sup }\left\vert g_{2}\left( \tau
,z\left( \tau \right) \right) \right\vert  \notag \\
&\leq &\sum\limits_{{i=1,i\neq p}}^{{n}}2^{1-\alpha _{i}}k_{i}h_{i}^{\alpha
_{i}}\varepsilon ^{i\alpha _{i}}+\varepsilon ^{n+1}L_{a}+2^{1-\alpha
_{p}}k_{p}L_{d}^{\alpha _{p}}  \notag \\
&\leq &\varepsilon ^{\rho }\delta _{0}+2^{1-\alpha _{p}}k_{p}L_{d}^{\alpha
_{p}}
\end{eqnarray}%
where $\delta _{0}=\sum\limits_{{i=1,i\neq p}}^{{n}}2^{1-\alpha
_{i}}k_{i}h_{i}^{\alpha _{i}}+L_{a}$, and

\begin{equation*}
\rho =\underset{i\in \left\{ 1,\cdots ,n\right\} ,i\neq p}{\min }\left\{
\min \{n+1,i\alpha _{i}\}\right\} =\alpha _{1}
\end{equation*}

From Proposition 8.1 in [21], Theorem 5.2 in [22] and Equation (52), there
exist positive constants $\mu $ and $\Gamma \left( z\left( 0\right) \right) $%
, such that

\begin{eqnarray}
\left\Vert z\left( \tau \right) \right\Vert &\leq &\mu \delta ^{\gamma }\leq
\mu (\varepsilon ^{\alpha _{1}}\delta _{0}+2^{1-\alpha
_{p}}k_{p}L_{d}^{\alpha _{p}})^{\gamma },  \notag \\
\forall \tau &\in &[\Gamma \left( z\left( 0\right) \right) ,\infty )
\end{eqnarray}

Therefore, from coordinate transformation (48), we obtain

\begin{eqnarray}
\Vert \left[
\begin{array}{ccc}
\varepsilon e_{1} & \cdots & \varepsilon ^{n}e_{n}%
\end{array}%
\right] \Vert &\leq &\mu (\varepsilon ^{\alpha _{1}}\delta _{0}+2^{1-\alpha
_{p}}k_{p}L_{d}^{\alpha _{p}})^{\gamma },  \notag \\
\forall t &\in &[\varepsilon \Gamma \left( {\Xi (\varepsilon )e}\left( {0}%
\right) \right) ,\infty )
\end{eqnarray}

Thus, the following inequality holds:

\begin{equation}
\left\vert e_{i}\right\vert \leq L(\delta _{di})^{\gamma },i=1,\cdots
,n,\forall t\in \lbrack \varepsilon \Gamma \left( {\Xi (\varepsilon )e}%
\left( {0}\right) \right) ,\infty )
\end{equation}%
where $L=\mu \delta _{0}^{\gamma }$, $\delta _{di}=\varepsilon ^{\alpha _{1}-%
\frac{i}{\gamma }}+\frac{2^{1-\alpha _{p}}}{\delta _{0}}k_{p}L_{d}^{\alpha
_{p}}\varepsilon ^{-\frac{i}{\gamma }}$, $i=1,\cdots ,n$. If $\varepsilon
\in \left( 0,1\right) $ and $L_{d}<\left( \frac{1-\varepsilon ^{\alpha _{1}}%
}{2^{1-\alpha _{p}}k_{p}}\delta _{0}\right) ^{\frac{1}{\alpha _{p}}}$, then

\begin{equation}
0<\varepsilon ^{\alpha _{1}}+\frac{2^{1-\alpha _{p}}}{\delta _{0}}%
k_{p}L_{d}^{\alpha _{p}}<1
\end{equation}

Furthermore, from Theorem 4.3 in [22], $\theta $\ can be chosen to be
arbitrarily small. Hence, the requirement that $\theta $\ lies on

\begin{equation}
\theta \in \left( 0,\min \left\{ \frac{1}{\frac{(n+1)\log \varepsilon }{\log
(\varepsilon ^{\alpha _{1}}+\frac{2^{1-\alpha _{p}}}{\delta _{0}}%
k_{p}L_{d}^{\alpha _{p}})}+1},\frac{1}{2}\right\} \right)
\end{equation}%
is not restrictive. Accordingly, we can obtain $\gamma =(1-\theta )/\theta
>\max \left\{ \frac{(n+1)\log \varepsilon }{\log (\varepsilon ^{\alpha _{1}}+%
\frac{2^{1-\alpha _{p}}}{\delta _{0}}k_{p}L_{d}^{\alpha _{p}})},1\right\} $.
Therefore,

\begin{equation}
\gamma \log (\varepsilon ^{\alpha _{1}}+\frac{2^{1-\alpha _{p}}}{\delta _{0}}%
k_{p}L_{d}^{\alpha _{p}})<(n+1)\log \varepsilon
\end{equation}%
i.e.,

\begin{equation}
\varepsilon ^{\alpha _{1}}+\frac{2^{1-\alpha _{p}}}{\delta _{0}}%
k_{p}L_{d}^{\alpha _{p}}<\varepsilon ^{\frac{n+1}{\gamma }}
\end{equation}

Therefore, from $\varepsilon \in \left( 0,1\right) $ and $\gamma >n+1$, we
can obtain $\varepsilon ^{\frac{n+1}{\gamma }}<\varepsilon ^{\frac{i}{\gamma
}},i=1,\cdots ,n$. Then

\begin{equation}
\delta _{di}=\varepsilon ^{\alpha _{1}-\frac{i}{\gamma }}+\frac{2^{1-\alpha
_{p}}}{\delta _{0}}k_{p}L_{d}^{\alpha _{p}}\varepsilon ^{-\frac{i}{\gamma }%
}<1,i=1,\cdots ,n
\end{equation}

The choice of $\theta $ leads to $\gamma >1$ in (55) which implies that for $%
\delta _{di}\in (0,1)$, the ultimate bound (55) on the estimation error is
of higher order than the perturbation. Consequently, the presented
integral-derivative observer leads to perform rejection of low-level
persistent disturbances. This concludes the proof. $\blacksquare $

\bigskip

\textbf{5. Computational analysis and simulations}

In this section, simulation results are presented in order to observe the
performances of the proposed integral-derivative observer. We consider the
simulations of the following control systems: 1) Integral-derivative
observer for a input signal; 2) PID control based on integral-derivative
observer for a second-order system.

Here, the stochastic non-white noise $\delta (t)$ is selected, and the mean
value of the noise is not equal to zero (See the noise in Figure 1(a)). The
non-white noise consists of following two signals: Random number with
Mean=0, Variance=0.01, Initial speed=0, and Sample time=0; Pulses with
Amplitude=0.5, Period=1s, Pulse width=1, and Phase delay=0.

\bigskip

\emph{1) Integral-derivative observer for a input signal with non-white noise%
}

For the integral-derivative observer

\begin{eqnarray}
\dot{x}_{1} &=&x_{2}  \notag \\
\dot{x}_{2} &=&x_{3}  \notag \\
\varepsilon ^{4}\dot{x}_{3} &=&-k_{1}\left\vert \varepsilon x_{1}\right\vert
^{\alpha _{1}}sign\left( x_{1}\right)  \notag \\
&&-k_{2}\left\vert x_{2}-a(t)\right\vert ^{\alpha _{2}}sign\left(
x_{2}-a(t)\right)  \notag \\
&&-k_{3}\left\vert \varepsilon ^{3}x_{3}\right\vert ^{\alpha _{3}}sign\left(
x_{3}\right)
\end{eqnarray}%
let the input signal be $a(t)=a_{0}(t)+\delta (t)$, where $a_{0}(t)=\cos t$
is the desired input signal, and $\delta (t)$ is the non-white noise.
Therefore, we obtain

\begin{equation}
\int_{0}^{t}a_{0}(\tau )d\tau =\sin t,\dot{a}_{0}(\tau )=-\sin t
\end{equation}

In the integral-derivative observer, $x_{2}$ tracks the reference signal $%
a_{0}(t)$; $x_{1}$ and $x_{3}$ estimate the 1-fold integral and first-order
derivative of $a_{0}(t)$, respectively. Observer parameters: $\varepsilon
=1/2$, $k_{1}=0.1$, $k_{2}=2$, $k_{3}=1$; $\alpha _{3}=0.8$, $\alpha _{2}=%
\frac{\alpha _{3}}{2-\alpha _{3}},\alpha _{1}=\frac{\alpha _{2}\alpha _{3}}{%
2\alpha _{3}-\alpha _{2}}$. Initial values of observer: $x_{1}\left(
0\right) =0$, $x_{2}\left( 0\right) =1$, $x_{3}\left( 0\right) =0$. Signal $%
a_{0}(t)$ tracking and the estimations of the first-order derivative and
1-fold integral are presented in Figure 1. Figure 1(a) provides signal $%
a_{0}\left( t\right) $ with stochastic noise. Figure 1(b) describes $%
a_{0}(t) $ tracking. Figures 1(c) and 1(d) present the estimations of the
first-order derivative and 1-fold integral, respectively. From the above
simulations, despite the intensive stochastic noise, the proposed
integral-derivative observer showed a very promising tracking ability and
robustness.

Furthermore, we compare the presented integral-derivative observer with the
integral operator in Matlab Simulink module (See Figures 1(d) and 1(e)). We
did the simulation in 3000 seconds, and no drift phenomenon happened for the
presented integral-derivative observer (See Figure 1(e)). However, an
obvious drift exists in the integral output by the integral operator in
Simulink module (See Figure 1(e)). In fact, in MATLAB, some numerical
methods are used to estimate signal integral (for example, the trapezoidal
rule, Simpson's rule). For the above numerical integrating methods, if white
noise exists in signal, the noise can be restrained sufficiently because of
integration. However, if stochastic noise (especially non-white noise)
exists in signal, and the average value of the noise is not equal to zero,
then such noise leads to the accumulation of additional drift in the
integrated signal. It is not guaranteed that the system is stable.

\bigskip

\emph{2) PID control based on integral-derivative observer for second-order
system}

The following second-order system is considered:

\begin{eqnarray}
\dot{z}_{1} &=&z_{2}  \notag \\
\dot{z}_{2} &=&u
\end{eqnarray}%
where, $z_{1}$ and $z_{2}$ are the states, $u$ is the control input. The
measurement output is

\begin{equation}
y=z_{1}+\delta (t)
\end{equation}%
where $\delta (t)$ is the bounded high-frequency non-white noise. The
second-order system is the equivalent or simplified model for many
mechanical systems, for instance, inverted pendulum control systems,
aircraft attitude control systems, \emph{et al}.

We are interested in designing a PID controller $u$ to force the system to
asymptotically track a given reference signal without the information of $%
z_{2}$ and $\int_{0}^{t}z_{1}(\tau )d\tau $.

Let the reference trajectory be ($z_{d}$, $\dot{z}_{d}$). The goal of
control is that

\begin{equation}
z_{1}\rightarrow z_{d},z_{2}\rightarrow \dot{z}_{d}
\end{equation}
as $t\rightarrow \infty $. For the reference trajectory ($z_{d}$, $\dot{z}%
_{d}$), let $e_{1}=z_{1}-z_{d}$ and $e_{2}=z_{2}-\dot{z}_{d}$. The system
error is

\begin{eqnarray}
\dot{e}_{1} &=&e_{2}  \notag \\
\dot{e}_{2} &=&u-\ddot{z}_{d}(t)
\end{eqnarray}

If $z_{1}$, $z_{2}$ and $\int_{0}^{t}z_{1}(\tau )d\tau $ are all known, the
PID controller can be designed as:

\begin{equation}
u=K_{P}e_{1}+K_{I}\int_{0}^{t}e_{1}(\tau )d\tau +K_{D}\dot{e}_{1}+\ddot{z}%
_{d}(t)
\end{equation}

Therefore, the closed-loop error system is

\begin{eqnarray}
\dot{e}_{1} &=&e_{2}  \notag \\
\dot{e}_{2} &=&K_{P}e_{1}+K_{I}\int_{0}^{t}e_{1}(\tau )d\tau +K_{D}\dot{e}%
_{1}(t)
\end{eqnarray}

Let

\begin{equation}
w_{1}=\int_{0}^{t}e_{1}(\tau )d\tau ,w_{2}=e_{1}(t),w_{3}=e_{2}(t)
\end{equation}

Therefore, it follows that

\begin{eqnarray}
\dot{w}_{1} &=&w_{2}  \notag \\
\dot{w}_{2} &=&w_{3}  \notag \\
\dot{w}_{3} &=&K_{I}w_{1}+K_{P}w_{2}+K_{D}w_{3}
\end{eqnarray}

The parameters $K_{I}$, $K_{P}$ and $K_{D}$ are selected such that $%
s^{3}+K_{D}s^{2}++K_{P}s+K_{I}$ is Hurwitz, then the closed-loop system is
stable.

However, $z_{2}$ and $\int_{0}^{t}z_{1}(\tau )d\tau $ are unknown, and the
non-white noise $\delta (t)$ exists in the measurement output $%
y=z_{1}+\delta (t)$. Here, the presented integral-derivative observer is
used to estimate these unknown variables from the measurement output $y$,
and the noise $\delta (t)$ is reduced sufficiently. The integral-derivative
observer is designed as

\begin{eqnarray}
\dot{x}_{1} &=&x_{2}  \notag \\
\dot{x}_{2} &=&x_{3}  \notag \\
\varepsilon ^{4}\dot{x}_{3} &=&-k_{1}\left\vert \varepsilon x_{1}\right\vert
^{\alpha _{1}}sign\left( x_{1}\right)  \notag \\
&&-k_{2}\left\vert x_{2}-y\right\vert ^{\alpha _{2}}sign\left( x_{2}-y\right)
\notag \\
&&-k_{3}\left\vert \varepsilon ^{3}x_{3}\right\vert ^{\alpha _{3}}sign\left(
x_{3}\right)
\end{eqnarray}%
where $x_{2}$ tracks the state $z_{1}$; $x_{1}$ and $x_{3}$ estimate the
integral and derivative of state $z_{1}$, respectively.

The PID controller is designed as

\begin{equation}
u=K_{P}\widehat{e}_{2}+K_{I}\widehat{e}_{1}+K_{D}\widehat{e}_{3}+\ddot{z}%
_{d}(t)
\end{equation}

where $\widehat{e}_{2}=x_{2}-z_{d}(\tau )$, $\widehat{e}_{1}=x_{1}-%
\int_{0}^{t}z_{d}(\tau )d\tau $, $\widehat{e}_{3}=x_{3}-\dot{z}_{d}(\tau )$.

Let the reference trajectory $(z_{d},\dot{z}_{d})=(\cos t,-\sin t)$.
Therefore, $\int_{0}^{t}z_{d}(\tau )d\tau =\sin t$ and $\ddot{z}_{d}=-\cos t$%
.

Observer parameters: $\varepsilon =1/3$, $k_{1}=0.1$, $k_{2}=2$, $k_{3}=1$, $%
\alpha _{3}=0.9$, $\alpha _{2}=\frac{\alpha _{3}}{2-\alpha _{3}},\alpha _{1}=%
\frac{\alpha _{2}\alpha _{3}}{2\alpha _{3}-\alpha _{2}}$; the initial value
of the second system is ($z_{1}\left( 0\right) =0.5$, $z_{2}\left( 0\right)
=-0.5$); the initial value of the observer is ($x_{1}\left( 0\right) =0$, $%
x_{2}\left( 0\right) =0.5$, $x_{3}\left( 0\right) =-0.5$); controller
parameters: $K_{P}=-2$, $K_{I}=-1$, $K_{D}=-1$.

Figure 2 shows the trajectory tracking and the estimations of derivative and
integral for the second-order system. Figure 2(a) describes the measurement
output $y$ and noise; Figure 2(b) describes the estimation and tracking of $%
z_{1}$; Figure 2(c) describes the estimation of $\int_{0}^{t}z_{1}(\tau
)d\tau $; Figure 2(d) describes the estimation and tracking of $z_{2}$;
Figure 2(e) presents the controller $u$. In the simulation above, though
stochastic noises exist in the measurement output, the estimations by the
presented integral-derivative observer and the control results by the
designed PID controller have satisfying qualities. Even in long-time
simulation, no drift phenomenon happen, and the estimations are accurate.
However, from Figure 2(c), an obvious drift exists in the integral output by
the integral operator in Simulink module. The integral algorithm can't
restrain the effect of stochastic noise (especially non-white noise). Such
noise leads to the accumulation of additional drift in the integrated signal.

\bigskip

\textbf{6. Conclusions}

In this paper, a nonlinear integral-derivative observer based on finite-time
stability is presented. The proposed integral-derivative observer can
estimate the integrals and derivatives of a signal synchronous. The
parameters selection is only required to be satisfied with Hurwitz
condition. Furthermore, the integral-derivative observer exhibits excellent
robustness and dynamical performance, and almost without drift phenomenon.

\bigskip

\textbf{Acknowledgements}

This research is supported in part by Australian Research Council (ARC)
Discovery (Grant nos. DP 0986814, DP 110104970), ARC linkage infrastructure,
Equipment and Facilities (Grant nos. LE 0347024, LE 0668508).

\begin{figure}[h]
\begin{center}
\includegraphics[width=2.80in]{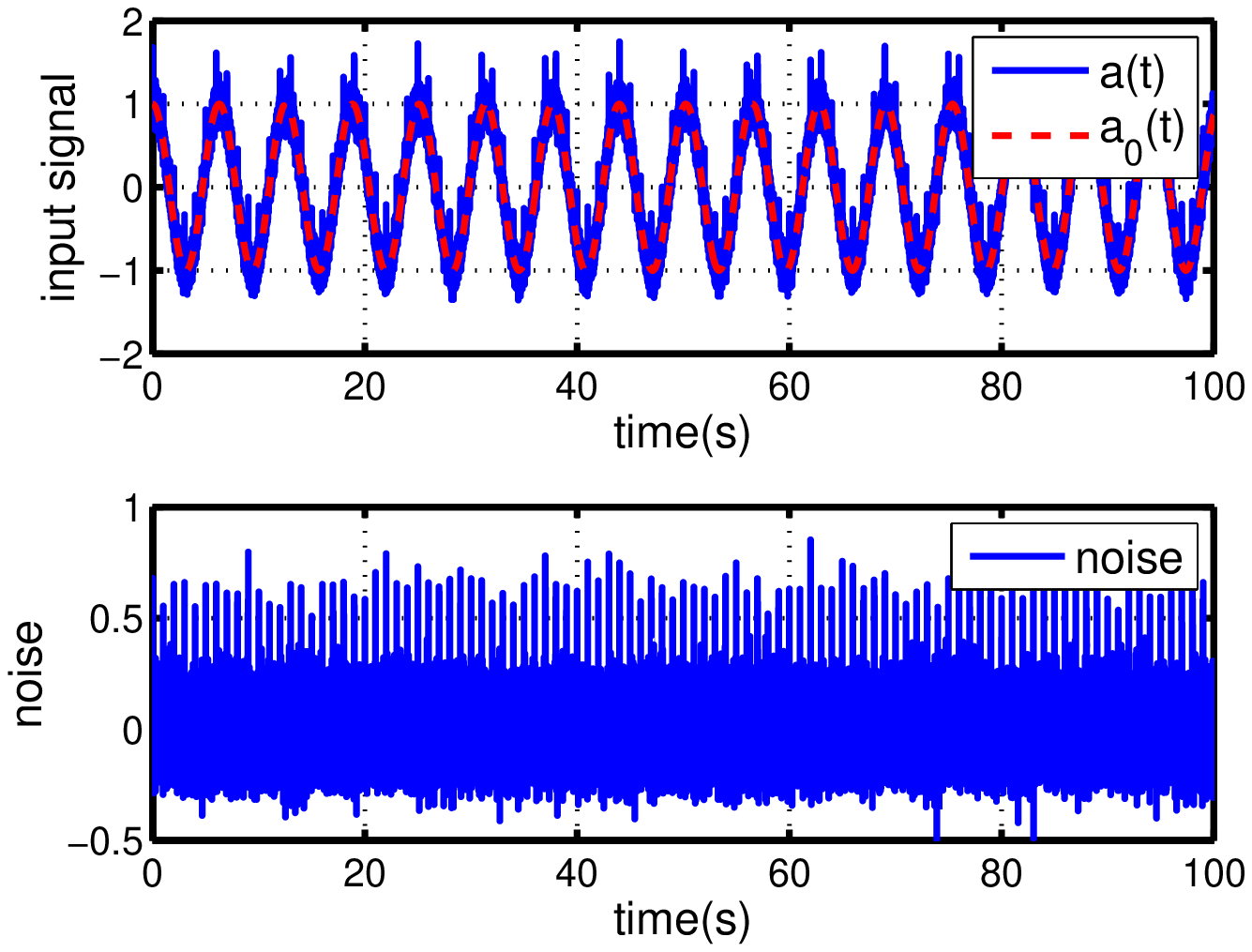}\\[0pt]
{1(a) Input signal\\[0pt]
\includegraphics[width=2.80in]{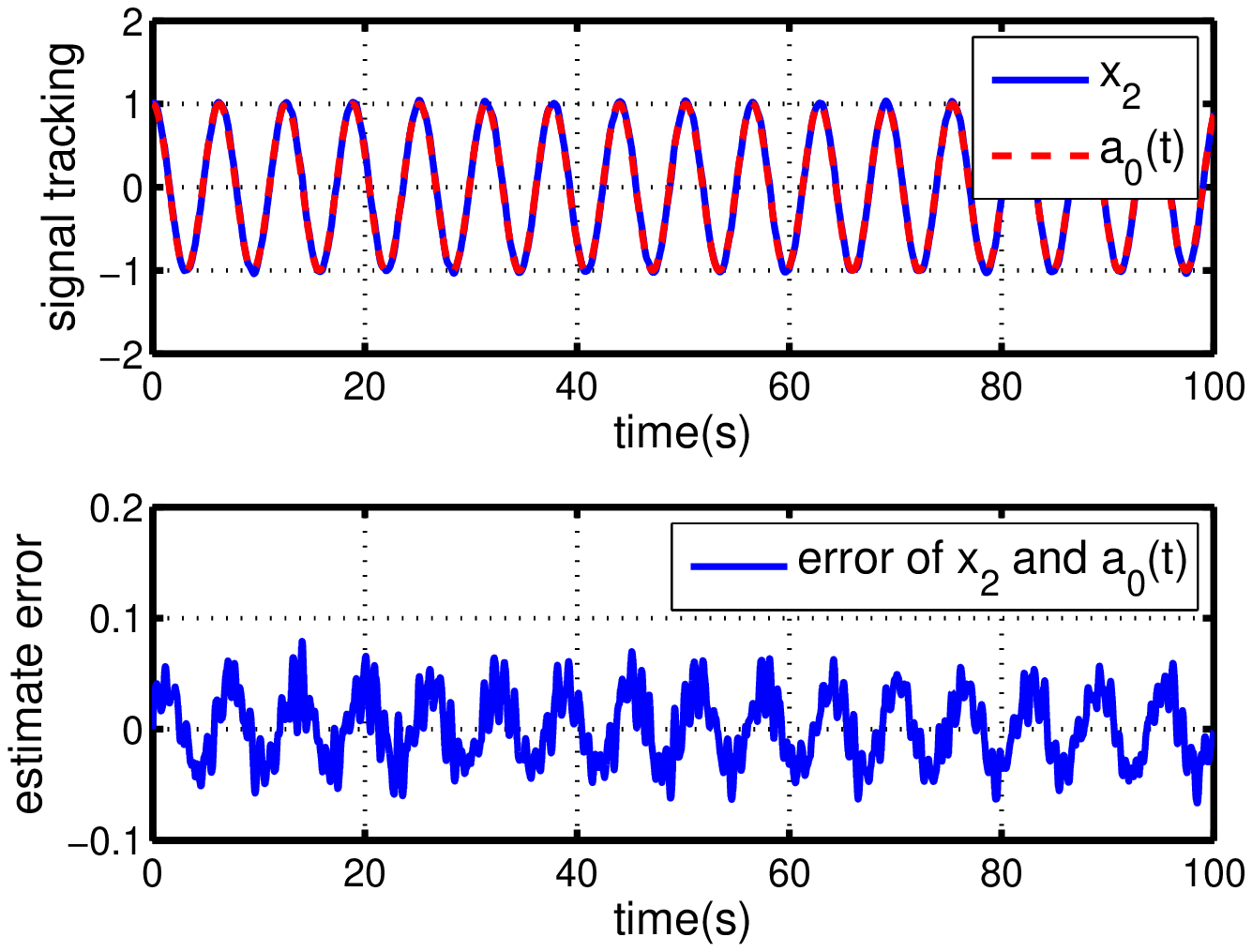}\\[0pt]
1(b) }$a_{0}(t)$\ tracking\\[0pt]
\includegraphics[width=2.80in]{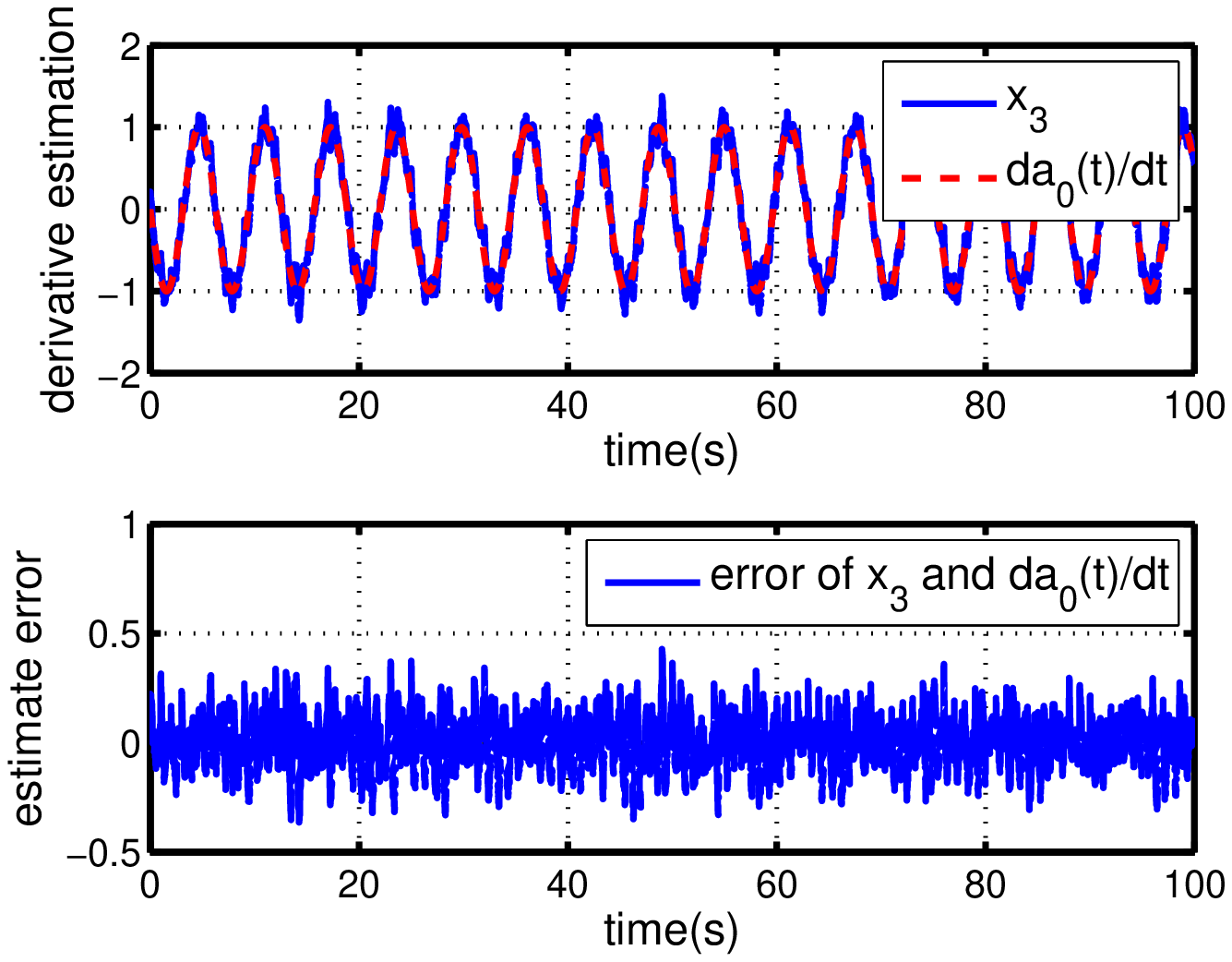}\\[0pt]
1(c) Derivative estimate\\[0pt]
\includegraphics[width=2.80in]{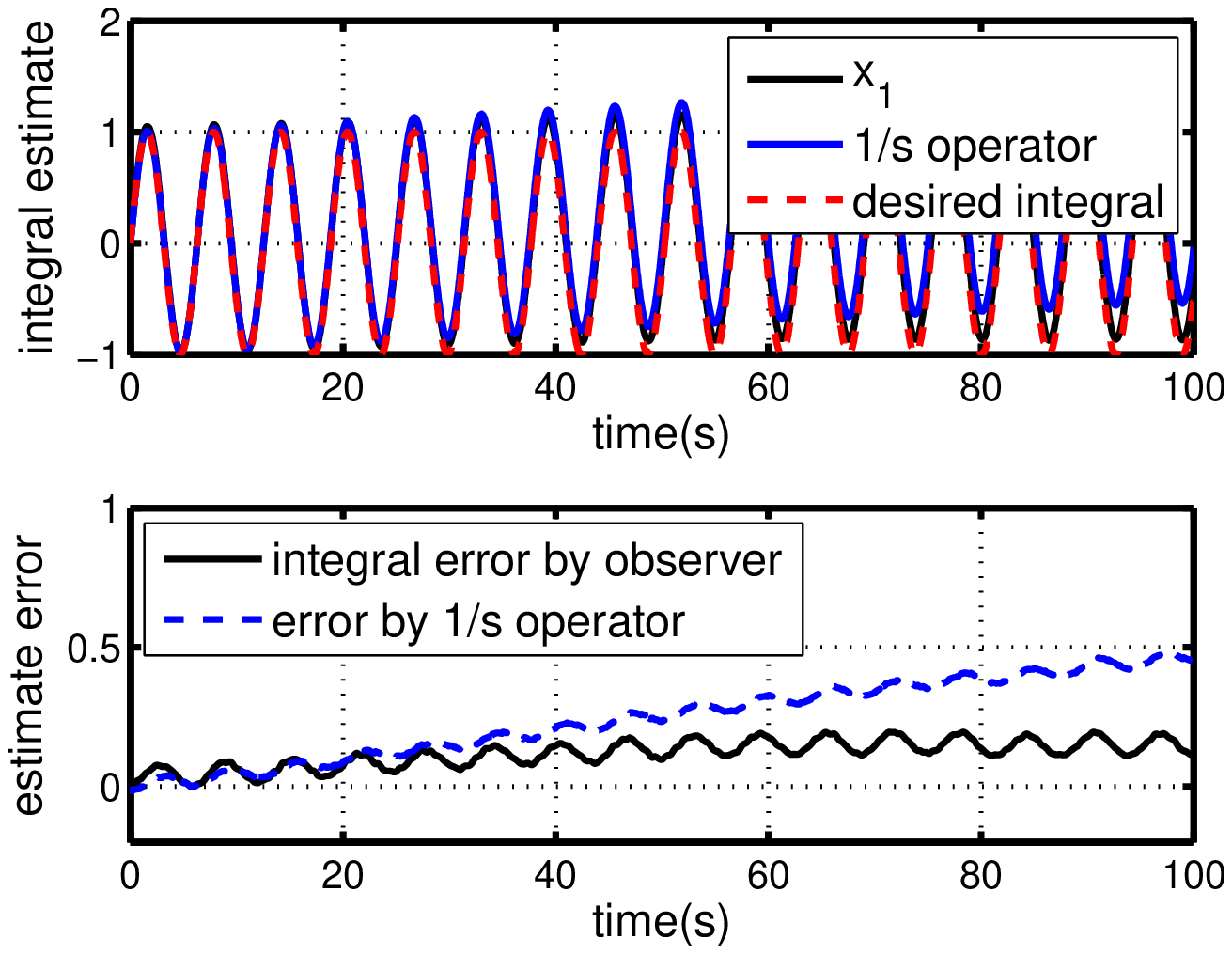}\\[0pt]
1(d) Integral estimate in 100s\\[0pt]
\end{center}
\end{figure}

\begin{figure}[h]
\begin{center}
\includegraphics[width=2.80in]{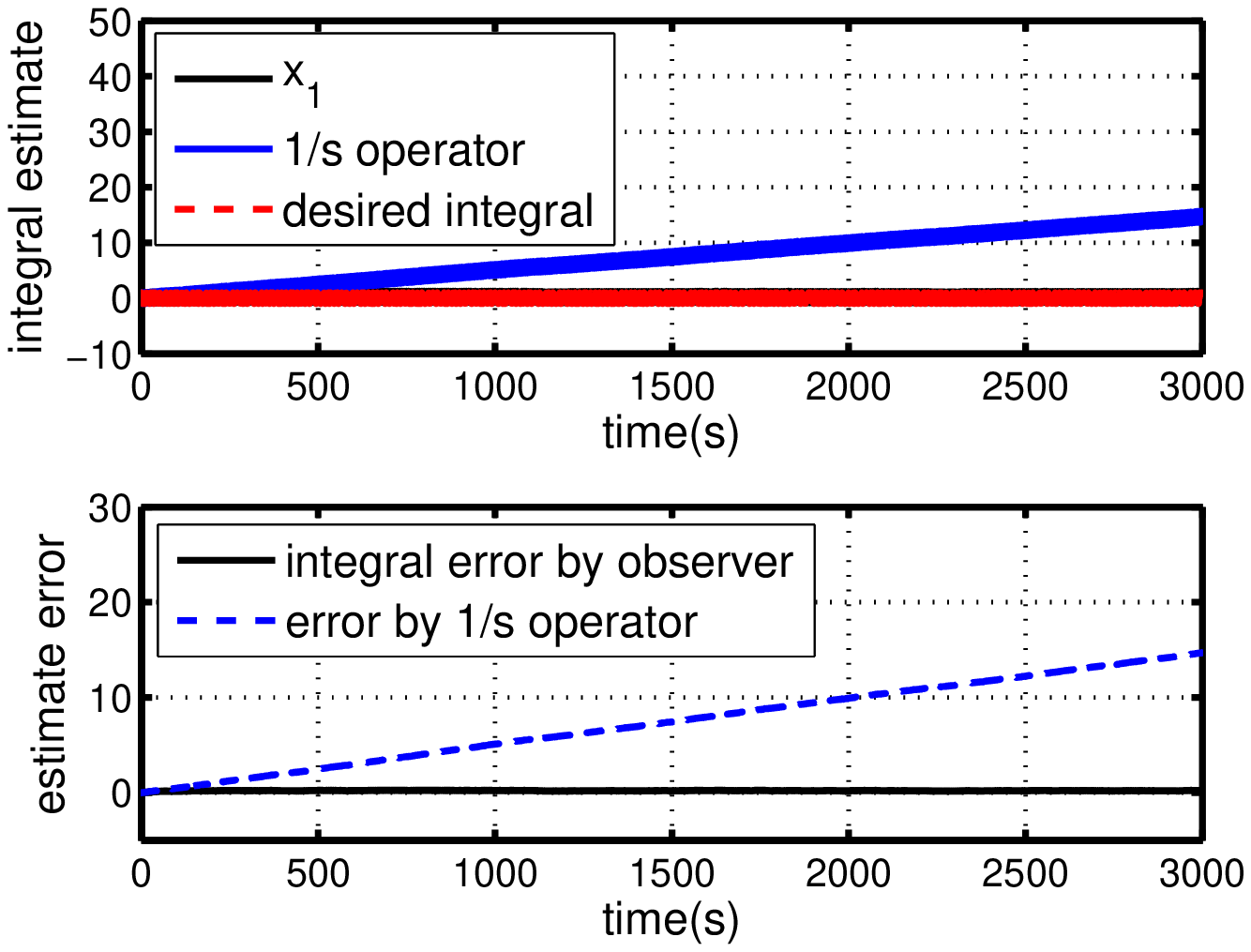}\\[0pt]
1(e) Integral estimate in 3000s\\[0pt]
{Figure 1 Integral-derivative observer}\\[0pt]
\includegraphics[width=2.80in]{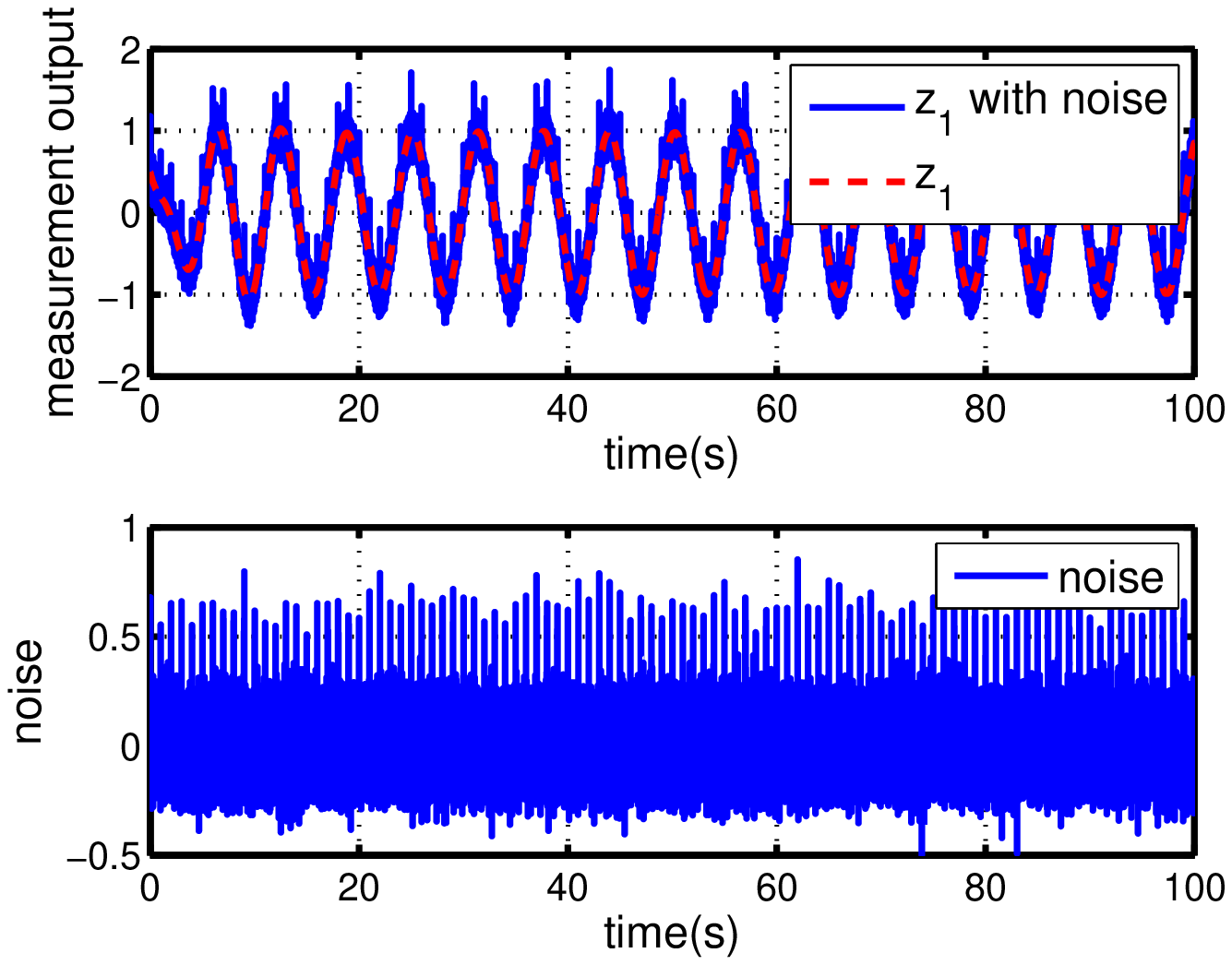}\\[0pt]
2(a) Measurement output\\[0pt]
\includegraphics[width=2.80in]{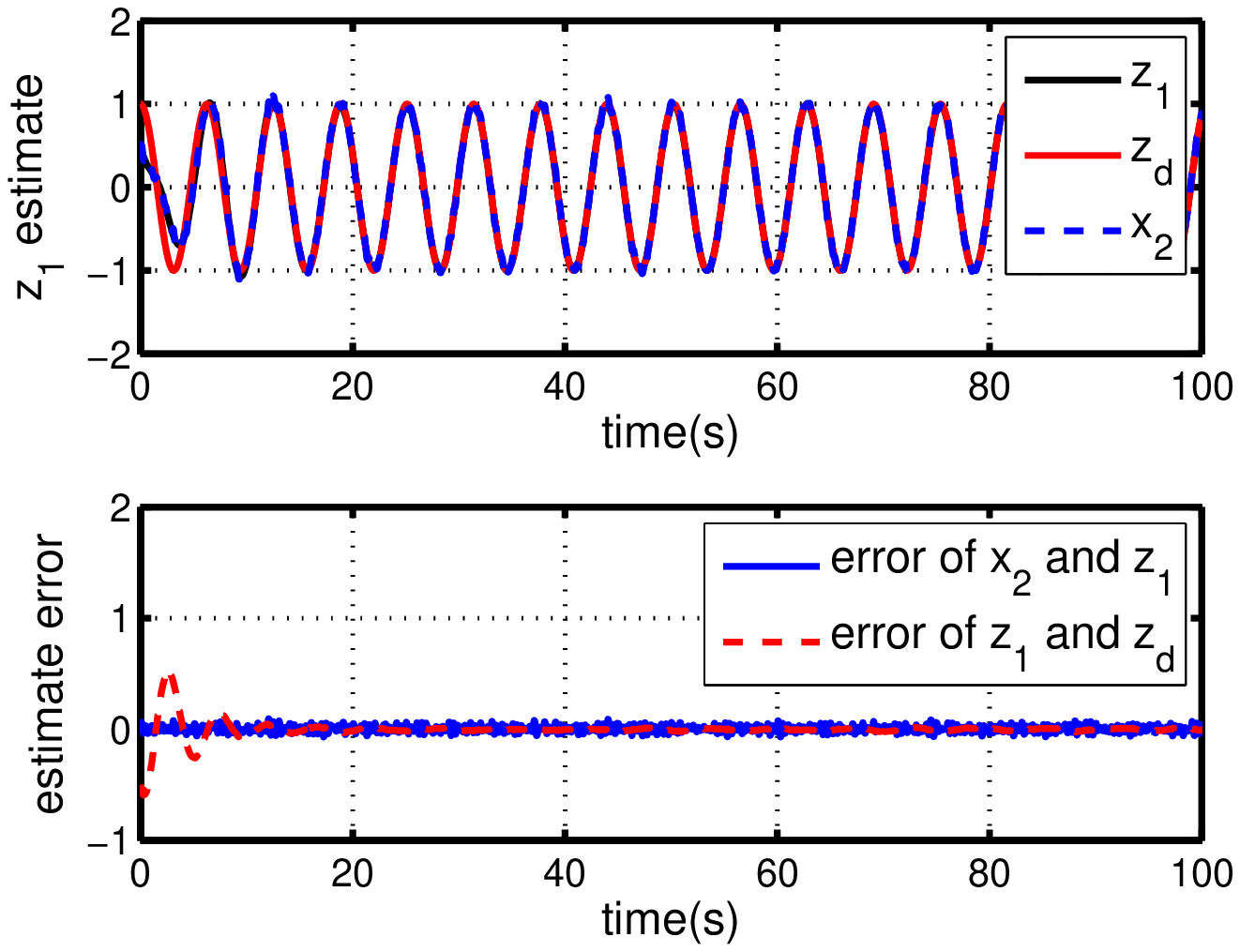}\\[0pt]
{2(b) }$z_{1}$ estimate\\[0pt]
\includegraphics[width=2.80in]{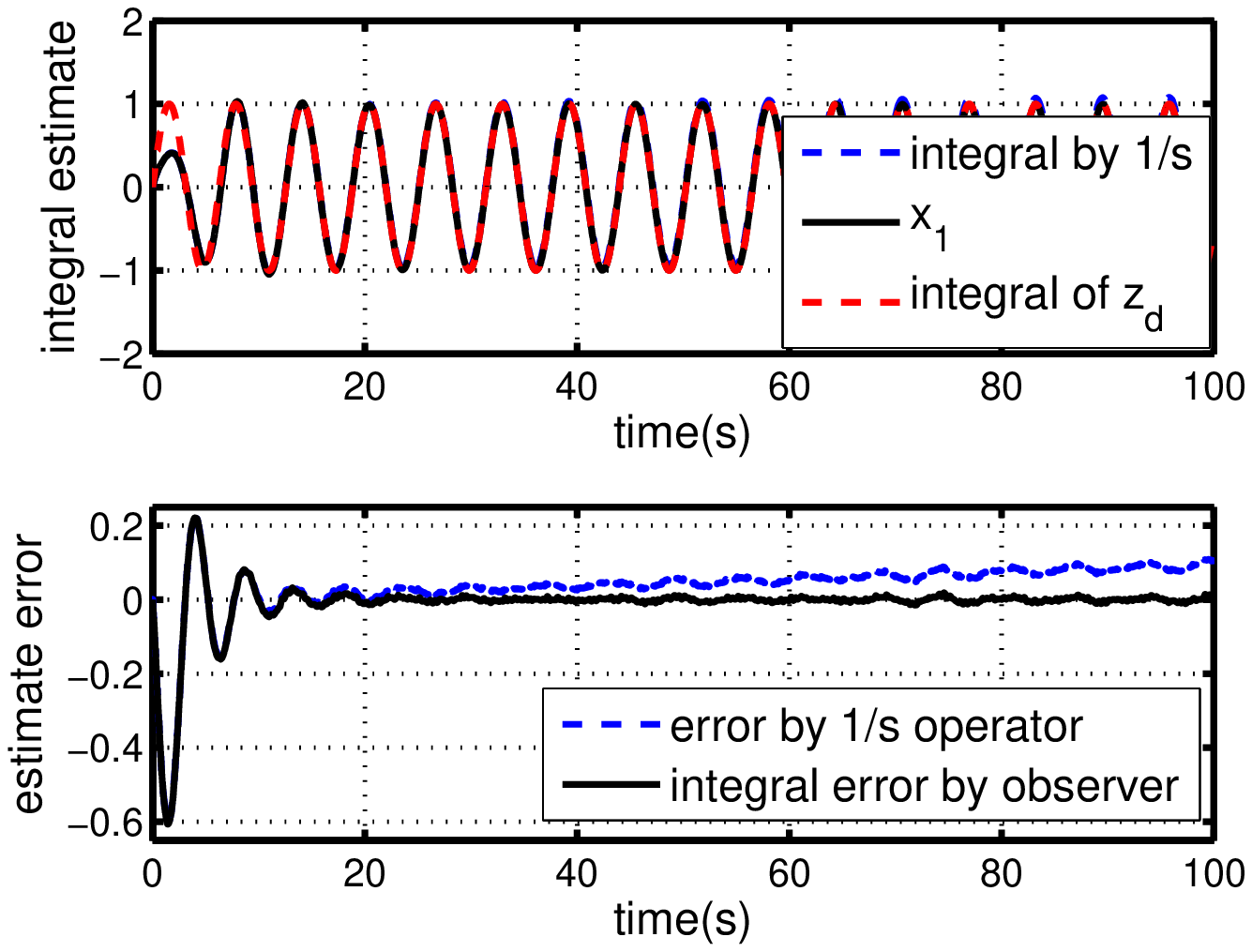}\\[0pt]
{2(c) }Estimate of integral of $z_{1}$%
\end{center}
\end{figure}

\begin{figure}[h]
\begin{center}
\includegraphics[width=2.80in]{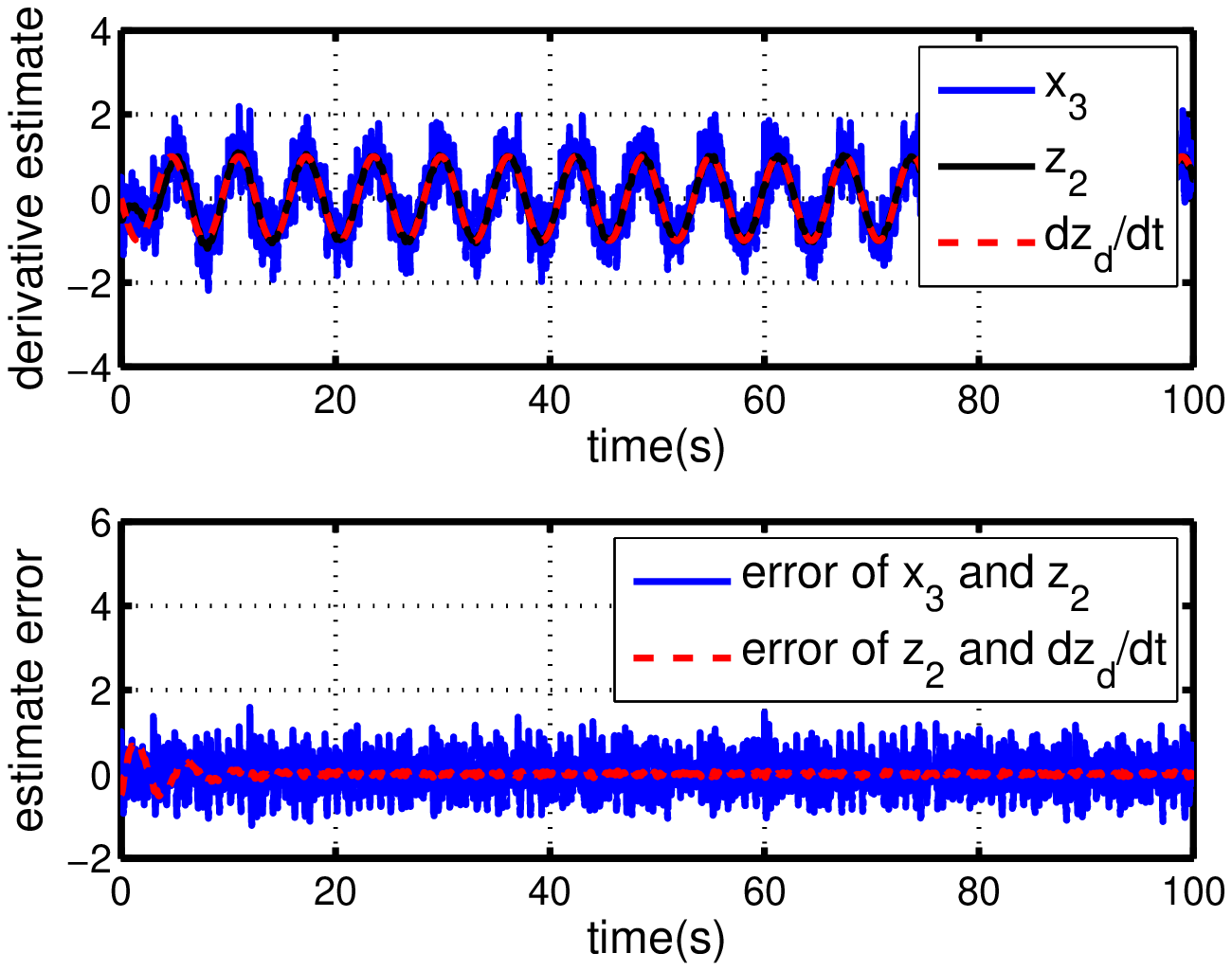}\\[0pt]
{2(d) }$z_{2}$ estimate{\\[0pt]
\includegraphics[width=2.80in]{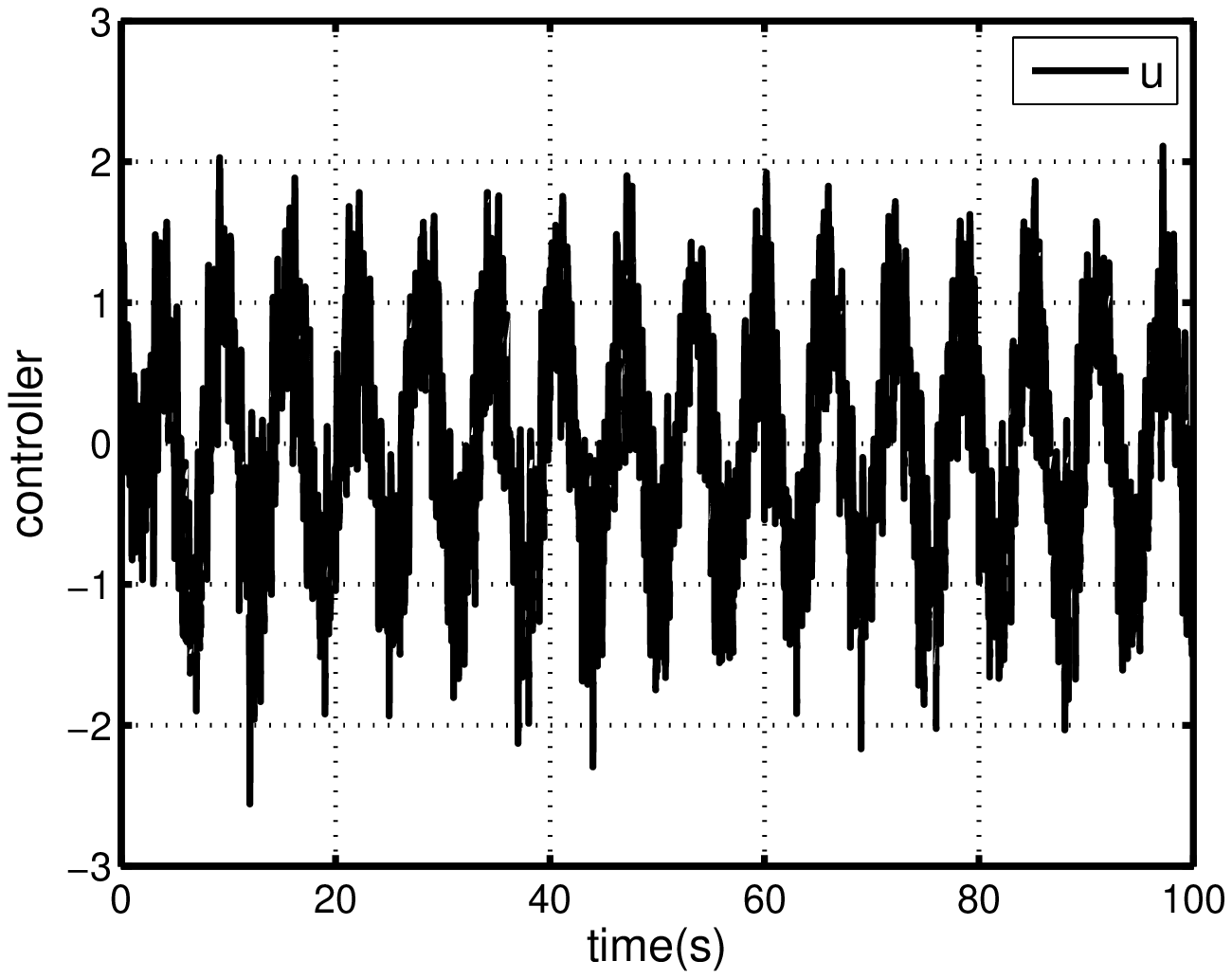}\\[0pt]
2(e) }Controller $u${\\[0pt]
Figure 2 PID control based on integral-derivative observer}
\end{center}
\end{figure}


\begin{thebibliography}{99}
\bibitem{} Takehira, T., Vinh, N.X., Kabamba, P.T.: Analytical solution of
missile terminal guidance. AIAA Guidance, Navigation, and Control
Conference, New Orleans, LA, United States, 11-13, 172-178 (1997)

\bibitem{} Drakunov, V.S., Ozguner, Dix, P., Ashrafi, B.: ABS control using
optimum search via sliding modes. IEEE Transactions on Control Systems
Technology, 3(1). 79-85 (1995)

\bibitem{} Tseng, C.C.: Digital integrator design using Simpson rule and
fractional delay filter. IEE Proceedings - Vision, Image and Signal
Processing, 153(1), 79-86 (2006)

\bibitem{} Tseng, C.C., Lee, S.L.: Digital IIR integrator design using
recursive Romberg integration rule and fractional sample delay. Signal
Processing, 88(9), 2222-2233 (2008)

\bibitem{} Ngo, N.Q.: A new approach for the design of wideband digital
integrator and differentiator. IEEE Transactions on Circuits and Systems II:
Express Briefs, 53(9), 936-940 (2006)

\bibitem{} Hodges, T., Nelson, P.A., Elliott, S.J.: The design of a
precision digital integrator for use in an active vibration control system.
Mechanical Systems and Signal Processing, 4(4), 345-353 (1990)

\bibitem{} Al-Alaoui, M.A.: A novel approach to designing a noninverting
integrator with built-in low frequency stability, high frequency
compensation, and high Q. IEEE Transactions on Instrumentation and
Measurement, 38(6), 1116-1121 (1989)

\bibitem{} Al-Alaoui, M.A.: Novel digital integrator and differentiator.
Electronics Letters, 29(4), 376-378 (1993)

\bibitem{} Al-Alaoui, M.A.: A Class of Second Order Integrators and Lowpass
Differentiators. IEEE Transactions on Circuits and Systems---I: Fundamental
Theory and Applications, 42(4), 220-223 (1995)

\bibitem{} Al-Alaoui, M.A.: Low-frequency differentiators and integrators
for biomedical and seismic signals, IEEE Transactions on Circuits and
Systems---I: Fundamental Theory and Applications, 48(8), 1006-1011 (2001)

\bibitem{} Al-Alaoui, M.A.: Class of digital integrators and
differentiators. IET Signal Process. 5(2), 251--260 (2011)

\bibitem{} Hahn, B.H., Valentine, D.T.: Essential MATLAB for Engineers and
Scientists, 4nd ed Elsevier Ltd. (2010)

\bibitem{} Charef, A., Sun, H.H., Tsao, Y.Y., Onaral, B.: Fractal system as
represented by singularity function, IEEE Transactions on Automatic Control,
37(9), 1465-1470 (1992)

\bibitem{} Chiaref, A.: Analogue realisation of fractional-order integrator,
differentiator and fractional $PI^{\lambda }D^{\mu }$ controller, IEE Proc.
- Control Theory Appl., 153(6), 714-720 (2006)

\bibitem{} Pei, S.C., Shyu, J.J.: Design of FIR Hilbert Transformers and
differentiators by eigenfilter. IEEE Trans. Acoust. Speech Signal Process.,
ASSP(37), 505-511 (1989)

\bibitem{} Atassi, A.N., Khalil, H.K.: Separation results for the
stabilization of nonlinear systems using different high-gain observer
designs. Systems and Control Letters, 39, 183-191 (2000)

\bibitem{} Levant, A.: High-order sliding modes, differentiation and
output-feedback control. International Journal of Control, 76(9/10), 924-941
(2003)

\bibitem{} Wang, X., Chen, Z., Yang, G.: Finite-time-convergent
differentiator based on singular perturbation technique. IEEE Transactions
on Automatic Control, 52(9), 1731-1737 (2007)

\bibitem{} Wang, X., Shirinzadeh, B.: Rapid-convergent nonlinear
differentiator. Mechanical Systems and Signal Processing, 28, 414-431 (2012)

\bibitem{} Efimov, D. V., Fridman, L.: A hybrid robust non-homogeneous
finite-time differentiator. IEEE Transactions on Automatic Control, 56(5),
1213-1219 (2011)

\bibitem{} Bhat, S.P., Bernstein, D.S.: Geometric homogeneity with
applications to finite-time stability. Mathematics of Control, Signals, and
Systems, 17, 101--127 (2005)

\bibitem{} Bhat, S.P., Bemstein, D.S.: Finite-time stability of continuous
autonomous systems. Siam J. Control Optim., 38(3), 751-766 (2000)

\bibitem{} Haimo, V.T.: Finite time controllers. Siam J. Control Optim.,
24(4), 760--771 (1986)

\bibitem{} Li, S., Du, H., Lin, X.: Finite-time consensus algorithm for
multi-agent systems with double-integrator dynamics. Automatica, 47,
1706-1712 (2011)

\bibitem{} Sun, H., Li, S., Sun, C.: Finite time integral sliding mode
control of hypersonic vehicles, Nonlinear Dynamics, 73(1-2), 229-244 (2013)

\bibitem{} Hu, Q., Li, B., Zhang, A.: Robust finite-time control allocation
in spacecraft attitude stabilization under actuator misalignment, Nonlinear
Dynamics, 73,(1-2), 53-71 (2013)

\bibitem{} Aghababa, M. P., Aghababa, H.P.: Chaos suppression of rotational
machine systems via finite-time control method, Nonlinear Dynamics, 69(4),
1881-1888 (2012)

\bibitem{} Guo, Z., Huang, L.: Global exponential convergence and global
convergence in finite time of non-autonomous discontinuous neural networks,
Nonlinear Dynamics, 58(1-2), 349-359 (2009)

\bibitem{} Isidori, A., Sastry, S.S., Kokotovic, P.V., Byrnes, C. I.:
Singular perturbed zero dynamics of nonlinear systems, IEEE Trans. Automat.
Contr., 37, 1625-1631 (1992)

\bibitem{} Lee, J.I., Ha, I.J.: A novel approach to control of
nonminimum-phase nonlinear systems, IEEE Trans. Automat. Contr., 47,
1480-1486 (2002).

\bibitem{} Kahlil, H.K.: Nonlinear systems, 3nd ed. Englewood Cliffs, New
Jerse: Prentice-Hall (2002)
\end{thebibliography}
\end{document}